\UseRawInputEncoding
\documentclass[11pt]{article}
\pdfoutput=1
\usepackage{amsmath,amssymb,amsthm,amsxtra,overpic,bbm,bm,epsfig,ulem,color,multirow}
\usepackage{afterpage}
\usepackage{multirow}
\usepackage{rotating}
\usepackage{braket}

\begin{document}

\begin{center}
{\Large \bf Simiplified textures of the seesaw model \\
 for the trimaximal neutrino mixings}
\end{center}

\vspace{0.05cm}

\begin{center}
{\bf Zhen-hua Zhao\footnote{Corresponding author: zhaozhenhua@lnnu.edu.cn}, Xin-Yu Zhao \footnote{Co-second author} and Hui-Chao Bao \footnote{Co-second author} } \\
{ Department of Physics, Liaoning Normal University, Dalian 116029, China }
\end{center}

\vspace{0.2cm}

\begin{abstract}
In the seesaw framework, in the basis of $M^{}_{\rm R}$ being diagonal, we explore the simplified textures of $M^{}_{\rm D}$ that can naturally yield the trimaximal neutrino mixings and their consequences for the neutrino parameters and leptogenesis.
We first formulate the generic textures of $M^{}_{\rm D}$ that can naturally yield the trimaximal mixings and then examine if their parameters can be further reduced. Our analysis is restricted to the simple but instructive scenario that there is only one phase parameter $\phi$ responsible for both the CP violating effects at low energies and leptogenesis. Our attention is paid to the textures of $M^{}_{\rm D}$ that possess some vanishing or equal elements. On the basis of these results, we further examine if $\phi$ can also take a particular value. The consequences of the phenomenologically-viable simplified textures for the neutrino parameters and leptogenesis are studied. A concrete flavor-symmetry model that can realize one representation of them is given.
\end{abstract}

\newpage

\section{Introduction}

The observation of the phenomenon of neutrino oscillations indicates that neutrinos are massive and the lepton flavors are mixed \cite{xing}. In the literature, the most popular and natural way of generating the tiny but non-zero neutrino masses is the type-I seesaw mechanism, in which three right-handed neutrinos $N^{}_I$ (for $I=1, 2, 3$) are added into the SM \cite{seesaw}. These newly introduced fields not only have the usual Yukawa couplings with the left-handed neutrinos (which constitute the Dirac neutrino mass matrix $M^{}_{\rm D}$ after the eletroweak symmetry breaking) but also have their own Majorana mass matrix $M^{}_{\rm R}$. Without loss of generality, we will work in the basis of $M^{}_{\rm R} = {\rm diag}(M^{}_1, M^{}_2, M^{}_3)$ with $M^{}_I$ being three right-handed neutrino masses.
The essence of the seesaw mechanism is to have $M^{}_I$ be much larger than the electroweak scale, yielding an effective Majorana mass matrix for the light neutrinos as
\begin{eqnarray}
M^{}_{\nu} = M^{}_{\rm D} M^{-1}_{\rm R} M^{T}_{\rm D} \; .
\label{1}
\end{eqnarray}
Then, in the basis where the flavor eigenstates of three charged leptons are identical with their mass eigenstates, the neutrino mixing matrix $U$ \cite{pmns} is to be identified as the unitary matrix for diagonalizing $M^{}_\nu$:
\begin{eqnarray}
U^\dagger M^{}_\nu U^* =  {\rm diag}(m^{}_1, m^{}_2, m^{}_3)  \;,
\label{2}
\end{eqnarray}
with $m^{}_i$ being three light neutrino masses.
In the standard parametrization, $U$ is expressed in terms of three mixing angles $\theta^{}_{ij}$ (for $ij=12, 13, 23$), one Dirac CP phase $\delta$ and two Majorana CP phases $\rho$ and $\sigma$ as
\begin{eqnarray}
U  =
\left( \begin{matrix}
c^{}_{12} c^{}_{13} & s^{}_{12} c^{}_{13} & s^{}_{13} e^{-{\rm i} \delta} \cr
-s^{}_{12} c^{}_{23} - c^{}_{12} s^{}_{23} s^{}_{13} e^{{\rm i} \delta}
& c^{}_{12} c^{}_{23} - s^{}_{12} s^{}_{23} s^{}_{13} e^{{\rm i} \delta}  & s^{}_{23} c^{}_{13} \cr
s^{}_{12} s^{}_{23} - c^{}_{12} c^{}_{23} s^{}_{13} e^{{\rm i} \delta}
& -c^{}_{12} s^{}_{23} - s^{}_{12} c^{}_{23} s^{}_{13} e^{{\rm i} \delta} & c^{}_{23}c^{}_{13}
\end{matrix} \right) \left( \begin{matrix}
e^{{\rm i}\rho} &  & \cr
& e^{{\rm i}\sigma}  & \cr
&  & 1
\end{matrix} \right) \;,
\label{3}
\end{eqnarray}
where the abbreviations $c^{}_{ij} = \cos \theta^{}_{ij}$ and $s^{}_{ij} = \sin \theta^{}_{ij}$ have been used.

Neutrino oscillations are sensitive to six neutrino parameters: three mixing angles, two independent neutrino mass squared differences $\Delta m^2_{ij} \equiv m^2_i - m^2_j$ (for $ij =21, 31$), and $\delta$. Several groups have performed global analyses of the existing neutrino oscillation data to extract the values of these parameters \cite{global,global2}. For definiteness, we will use the results in Ref.~\cite{global} (see Table~1) as reference values in the following numerical calculations. Note that the sign of $\Delta m^2_{31}$ remains undetermined, thus allowing for two possible neutrino mass orderings: the normal ordering (NO) $m^{}_1 < m^{}_2 < m^{}_3$ and inverted ordering (IO) $m^{}_3 < m^{}_1 < m^{}_2$. In comparison, neutrino oscillations have nothing to do with the absolute values of neutrino masses and Majorana CP phases. In order to extract or constrain their values, one has to resort to some non-oscillatory experiments such as the neutrino-less double beta decay experiments \cite{0nbb}. Unfortunately, such experiments have not yet placed any lower constraint on the lightest neutrino mass, nor any constraint on the Majorana CP phases.

\begin{table}\centering
  \begin{footnotesize}
    \begin{tabular}{c|cc|cc}
     \hline\hline
      & \multicolumn{2}{c|}{Normal Ordering}
      & \multicolumn{2}{c}{Inverted Ordering }
      \\
      \cline{2-5}
      & bf $\pm 1\sigma$ & $3\sigma$ range
      & bf $\pm 1\sigma$ & $3\sigma$ range
      \\
      \cline{1-5}
      \rule{0pt}{4mm}\ignorespaces
       $\sin^2\theta^{}_{12}$
      & $0.318_{-0.016}^{+0.016}$ & $0.271 \to 0.370$
      & $0.318_{-0.016}^{+0.016}$ & $0.271 \to 0.370$
      \\[1mm]
       $\sin^2\theta^{}_{23}$
      & $0.566_{-0.022}^{+0.016}$ & $0.441 \to 0.609$
      & $0.566_{-0.023}^{+0.018}$ & $0.446 \to 0.609$
      \\[1mm]
       $\sin^2\theta^{}_{13}$
      & $0.02225_{-0.00078}^{+0.00055}$ & $0.02015 \to 0.02417$
      & $0.02250_{-0.00076}^{+0.00056}$ & $0.02039 \to 0.02441$
      \\[1mm]
       $\delta/\pi$
      & $1.20_{-0.14}^{+0.23}$ & $0.80 \to 2.00$
      & $1.54_{-0.13}^{+0.13}$ & $1.14 \to 1.90$
      \\[3mm]
       $\displaystyle \frac{\Delta m^2_{21}}{10^{-5}~{\rm eV}^2}$
      & $7.50_{-0.20}^{+0.22}$ & $6.94 \to 8.14$
      & $7.50_{-0.20}^{+0.22}$ & $6.94 \to 8.14$
      \\[3mm]
       $\displaystyle \frac{|\Delta m^2_{31}|}{10^{-3}~{\rm eV}^2}$
      & $2.56_{-0.04}^{+0.03}$ & $2.46 \to 2.65$
      & $2.46_{-0.03}^{+0.03}$ & $2.37 \to 2.55$
      \\[2mm]
      \hline\hline
    \end{tabular}
  \end{footnotesize}
  \caption{The best-fit values, 1$\sigma$ errors and 3$\sigma$ ranges of six neutrino
oscillation parameters extracted from a global analysis of the existing
neutrino oscillation data \cite{global}. }
\end{table}

From Table~1 we see that $\theta^{}_{12}$ and $\theta^{}_{23}$ are close to some special values: $\sin^2 \theta^{}_{12} \sim 1/3$ and $\sin^2 \theta^{}_{23} \sim 1/2$. Before the value of $\theta^{}_{13}$ was pinned down in 2012, the conjecture that it might be vanishingly small was ever very popular. For the ideal case of $\sin \theta^{}_{12} = 1/\sqrt{3}$, $\sin \theta^{}_{23} = 1/\sqrt{2}$ and $\theta^{}_{13} =0$ (referred to as the tribimaximal (TBM) mixing \cite{TB}), the neutrino mixing matrix can be described by a few simple numbers and their square roots
\begin{eqnarray}
U^{}_{\rm TBM}= \displaystyle \frac{1}{\sqrt 6} \left( \begin{array}{ccc}
2 & \sqrt{2} & 0 \cr
-1 & \sqrt{2} & \sqrt{3} \cr
1 & - \sqrt{2} & \sqrt{3} \cr
\end{array} \right)  \; .
\label{4}
\end{eqnarray}
Such a particular mixing might be suggestive of an underlying flavor symmetry in the lepton sector.
Along this direction, many flavor symmetries have been trialed to realize it \cite{FS}. But the actual relative largeness of $\theta^{}_{13}$ compels us to abandon or modify this mixing. An economical and predictive way out is to keep its first or second column unchanged while modifying the other two columns within the unitarity constraints, leading us to the first or second trimaximal (TM1 or TM2) mixing \cite{TM}. Such variants of $U^{}_{\rm TBM}$ can be obtained by multiplying it from the right-hand side by a complex (2,3) or (1,3) rotation matrix $U^{}_{23}$ or $U^{}_{13}$:
\begin{eqnarray}
U^{}_{\rm TM1}= U^{}_{\rm TBM} U^{}_{23} \hspace{0.5cm} {\rm with} \hspace{0.5cm} U^{}_{23} =\left( \begin{array}{ccc}
1 & 0  & 0 \cr
0 & \cos \theta & \sin \theta \hspace{0.03cm} e^{ -{\rm i} \varphi} \cr
0 & -\sin \theta \hspace{0.03cm} e^{ {\rm i} \varphi} & \cos \theta \cr
\end{array} \right) \;; \nonumber \\
U^{}_{\rm TM2}= U^{}_{\rm TBM} U^{}_{13} \hspace{0.5cm} {\rm with} \hspace{0.5cm} U^{}_{13} = \left( \begin{array}{ccc}
\cos \theta & 0  & \sin \theta \hspace{0.03cm} e^{ -{\rm i} \varphi} \cr
0 & 1 & 0 \cr
-\sin \theta \hspace{0.03cm} e^{ {\rm i} \varphi} & 0 & \cos \theta \cr
\end{array} \right) \; ,
\label{5}
\end{eqnarray}
where $\theta$ is a rotation angle and $\varphi$ a phase parameter.

By comparing $U^{}_{\rm TM1}$ and $U^{}_{\rm TM2}$ with the standard form of $U$ in Eq.~(\ref{3}), one can derive their consequences for the neutrino mixing parameters, among which we will use the following expressions of $s^2_{13}$ and $s^2_{23}$ to infer the values of $\theta$ and $\varphi$
\begin{eqnarray}
& & {\rm TM1}: \hspace{1cm} s^{2}_{13} = \frac{1}{3} \sin^2 \theta \; , \hspace{1cm}
s^{2}_{23} = \frac{1}{2} + \frac{\sqrt{6} \sin 2\theta \cos \varphi}{6 - 2 \sin^2 \theta}  \;; \nonumber \\
& & {\rm TM2}: \hspace{1cm} s^{2}_{13} = \frac{2}{3} \sin^2 \theta \; , \hspace{1cm}
s^{2}_{23} = \frac{1}{2} - \frac{ \sqrt{3} \sin 2\theta \cos \varphi}{6 - 4 \sin^2 \theta} \; .
\label{6}
\end{eqnarray}
Given the $3\sigma$ ranges of $s^2_{13}$ and $s^2_{23}$, $\theta$ and $|\varphi|$ are respectively constrained into the ranges 0.25---0.27 (0.17---0.19) and 0.36$\pi$---0.60$\pi$ (0.30$\pi$---0.99$\pi$) for the TM1 (TM2) mixing. Taking into account the relations in Eq.~(\ref{6}), one arrives at the following predictions for $\theta^{}_{12}$ and $\delta$ \cite{TM}
\begin{eqnarray}
&& {\rm TM1}: \hspace{1cm} s^{2}_{12} =  \frac{ 1}{3} -  \frac{2 s^{2}_{13}}{3 - 3s^{2}_{13}} \;, \hspace{1cm} \tan{2\theta^{}_{23}} \cos \delta = - \frac{1-5 s^2_{13}}{2 \sqrt{2} s^{}_{13} \sqrt{1- 3 s^2_{13}} } \;;
\nonumber \\
&& {\rm TM2}: \hspace{1cm} s^{2}_{12} = \frac{ 1}{3} + \frac{s^{2}_{13}}{3 - 3s^{2}_{13}} \;, \hspace{1cm} \tan{2\theta^{}_{23}} \cos \delta = \frac{1-2 s^2_{13}}{s^{}_{13} \sqrt{2- 3 s^2_{13}} } \; .
\label{7}
\end{eqnarray}
At the $3\sigma$ level, $s^2_{12}$ and $|\delta|$ are respectively constrained into the ranges 0.317---0.319 (0.340---0.342) and 0.33$\pi$---0.59$\pi$ (0.31$\pi$---1.00$\pi$) for the TM1 (TM2) mixing, which are in good agreement with the experimental results.

Due to their simple structures and phenomenologically-appealing consequences, the trimaximal mixings have attracted a lot of attention after the observation of a relatively large $\theta^{}_{13}$ \cite{TM2}-\cite{Tanimoto}. However, they are not restrictive enough so that not predictive enough. Therefore, in the literature many attempts to combine them with other constraints on the neutrino mass matrix have been made. For example, in Ref.~\cite{TMmutau} the authors have studied the combination of the trimaximal mixings with the $\mu$-$\tau$ reflection symmetry \cite{MTR}.
In Ref.~\cite{TM0} the authors have studied if one element of the $M^{}_\nu$ that can yield the trimaximal mixings is phenomenologically allowed to be vanishing (i.e., the combination of the trimaximal mixings with texture zeros). But this study has only been performed at the level of $M^{}_\nu$.
In a series of works about the so-called littlest seesaw model \cite{LS, littlest}, in the minimal seesaw framework, the authors have studied the simplified textures of $M^{}_{\rm D}$ that can yield the TM1 mixing. But this study is only confined to the minimal seesaw framework and the TM1 mixing.

In this paper, in the general seesaw framework, we will explore the simplified textures of $M^{}_{\rm D}$
that can naturally yield the trimaximal (including both the TM1 and TM2) mixings and study their consequences for the neutrino parameters and leptogenesis. From the simplicity viewpoint, our attention will be paid to the textures of $M^{}_{\rm D}$ that possess some vanishing or equal elements. But we will only focus on those that can find a simple symmetry justification.

The rest part of this paper is organized as follows. In the next section, we will first formulate the generic textures of $M^{}_{\rm D}$ that can naturally yield the trimaximal mixings and discuss how to realize them by slightly modifying the flavor-symmetry models for realizing the TBM mixing. Then, we will examine if their parameters can be further reduced, giving more simplified textures of them. The consequences of the phenomenologically-viable simplified textures of $M^{}_{\rm D}$ for the neutrino parameters and leptogenesis will be studied.
The studies for the TM1 and TM2 mixings will be performed in sections~3 and 4, respectively.
In section 5, a concrete flavor-symmetry model that can realize one representation of the obtained simplified textures of $M^{}_{\rm D}$ is given.
In section 6, we will discuss the impacts of the renormalization group running effects on our results.
Finally, our main results will be summarized in the last section.

\section{Generic textures of $M^{}_{\rm D}$ for the trimaximal mixings}

In this section, we first formulate the generic textures of $M^{}_{\rm D}$ that can naturally yield the trimaximal mixings. Let us parameterize the most generic $M^{}_{\rm D}$ as
\begin{eqnarray}
M^{}_{\rm D} = \left( \begin{array}{ccc} a^{}_{1} \sqrt{M^{}_1} & b^{}_{1} \sqrt{M^{}_2} & c^{}_1 \sqrt{M^{}_3} \cr
a^{}_{2} \sqrt{M^{}_1} & b^{}_{2} \sqrt{M^{}_2} & c^{}_2 \sqrt{M^{}_3} \cr a^{}_{3} \sqrt{M^{}_1} & b^{}_{3} \sqrt{M^{}_2} & c^{}_3 \sqrt{M^{}_3} \end{array} \right) \;,
\label{2.1}
\end{eqnarray}
where $a^{}_i$, $b^{}_i$ and $c^{}_i$ are generally complex parameters.
In terms of the QR parametrization, such an $M^{}_{\rm D}$ can be decomposed into $M^{}_{\rm D} = U^{}_{\rm L} \Delta$ \cite{QR}. Here $U^{}_{\rm L}$ is a unitary matrix as
\begin{eqnarray}
U^{}_{\rm L} = \left( \begin{array}{ccc}
\vspace{0.2cm}
\displaystyle \frac{a^{}_1}{ |\vec{a}| } & \displaystyle \frac{ b^{}_1 - \displaystyle \frac{\vec a \cdot \vec b}{|\vec{a}|^2} a^{}_1 } { \sqrt{ |\vec{b}|^2 - \displaystyle \frac{| \vec a \cdot \vec b|^2}{|\vec{a}|^2} } }  & \displaystyle \frac{ c^{}_1 - \displaystyle \frac{\vec a \cdot \vec c }{|\vec{a}|^2} a^{}_1 - \displaystyle \frac{|\vec{a}|^2 \vec{b} \cdot \vec c - \vec{b} \cdot \vec a \ \vec{a} \cdot \vec c  }{ |\vec{a}|^2  |\vec{b}|^2 - |\vec{a} \cdot \vec b|^2 }  \left( b^{}_1 - \displaystyle \frac{\vec{a} \cdot \vec b}{|\vec{a}|^2} a^{}_1 \right) } { \sqrt{ |\vec{c}|^2 - \displaystyle \frac{| \vec{a} \cdot \vec c|^2}{|\vec{a}|^2} -  \displaystyle \frac{ \left| |\vec{a}|^2  \vec{b} \cdot \vec c - \vec{b} \cdot \vec a \ \vec{a} \cdot \vec c \right|^2 }{|\vec{a}|^2 \left( |\vec{a}|^2  |\vec{b}|^2 - | \vec{a} \cdot \vec b|^2  \right) } } } \cr
\vspace{0.2cm}
\displaystyle \frac{a^{}_2}{ |\vec{a}| }  & \displaystyle \frac{ b^{}_2 - \displaystyle \frac{\vec a \cdot \vec b}{|\vec{a}|^2} a^{}_2 } { \sqrt{ |\vec{b}|^2 - \displaystyle \frac{| \vec a \cdot \vec b|^2}{|\vec{a}|^2} } }   & \displaystyle \frac{ c^{}_2 - \displaystyle \frac{\vec a \cdot \vec c }{|\vec{a}|^2} a^{}_2 - \displaystyle \frac{|\vec{a}|^2 \vec{b} \cdot \vec c - \vec{b} \cdot \vec a \ \vec{a} \cdot \vec c  }{ |\vec{a}|^2  |\vec{b}|^2 - |\vec{a} \cdot \vec b|^2 }  \left( b^{}_2 - \displaystyle \frac{\vec{a} \cdot \vec b}{|\vec{a}|^2} a^{}_2 \right) } { \sqrt{ |\vec{c}|^2 - \displaystyle \frac{| \vec{a} \cdot \vec c|^2}{|\vec{a}|^2} -  \displaystyle \frac{ \left| |\vec{a}|^2  \vec{b} \cdot \vec c - \vec{b} \cdot \vec a \ \vec{a} \cdot \vec c \right|^2 }{|\vec{a}|^2 \left( |\vec{a}|^2  |\vec{b}|^2 - | \vec{a} \cdot \vec b|^2  \right) } } } \cr
\displaystyle \frac{a^{}_3}{ |\vec{a}| } & \displaystyle \frac{ b^{}_3 - \displaystyle \frac{ \vec a \cdot \vec b }{|\vec{a}|^2} a^{}_3 } { \sqrt{ |\vec{b}|^2 - \displaystyle \frac{| \vec a \cdot \vec b|^2}{|\vec{a}|^2} } }   & \displaystyle \frac{ c^{}_3 - \displaystyle \frac{\vec a \cdot \vec c }{|\vec{a}|^2} a^{}_3 - \displaystyle \frac{|\vec{a}|^2 \vec{b} \cdot \vec c - \vec{b} \cdot \vec a \ \vec{a} \cdot \vec c  }{ |\vec{a}|^2  |\vec{b}|^2 - |\vec{a} \cdot \vec b|^2 }  \left( b^{}_3 - \displaystyle \frac{\vec{a} \cdot \vec b}{|\vec{a}|^2} a^{}_3 \right) } { \sqrt{ |\vec{c}|^2 - \displaystyle \frac{| \vec{a} \cdot \vec c|^2}{|\vec{a}|^2} -  \displaystyle \frac{ \left| |\vec{a}|^2  \vec{b} \cdot \vec c - \vec{b} \cdot \vec a \ \vec{a} \cdot \vec c \right|^2 }{|\vec{a}|^2 \left( |\vec{a}|^2  |\vec{b}|^2 - | \vec{a} \cdot \vec b|^2  \right) } } }
\end{array} \right) \;,
\label{2.2}
\end{eqnarray}
with $|\vec a| \equiv \sqrt{ |a^{}_1|^2 + |a^{}_2|^2 + |a^{}_3|^2 }$ and
$\vec a \cdot \vec b \equiv a^*_1 b^{}_1 + a^*_2 b^{}_2 + a^*_3 b^{}_3$ (and so on). On the other hand, $\Delta$ is a triangular matrix as
\begin{eqnarray}
\Delta  = \left( \begin{array}{ccc}
\vspace{0.2cm}
|\vec a| \sqrt{M^{}_1} & \displaystyle \frac{\vec a \cdot \vec b}{|\vec a|} \sqrt{M^{}_2}  &  \displaystyle \frac{\vec a \cdot \vec c}{|\vec a|} \sqrt{M^{}_3} \cr
\vspace{0.2cm}
0  &  \sqrt{  |\vec b|^2 - \displaystyle \frac{|\vec a \cdot \vec b|^2}{|\vec a|^2} } \sqrt{M^{}_2}   & \displaystyle \frac{ |\vec a|^2 \vec b \cdot \vec c -  \vec b \cdot \vec a \  \vec a \cdot \vec c } { |\vec a| \sqrt{|\vec a|^2 |\vec b|^2 - | \vec a \cdot \vec b |^2  } } \sqrt{M^{}_3} \cr
0 & 0 & \sqrt{ |\vec c|^2 - \displaystyle \frac{| \vec a \cdot \vec c|^2}{|\vec a|^2} -  \displaystyle \frac{ \left| |\vec a|^2 \vec b \cdot \vec c - \vec b \cdot \vec a \ \vec a \cdot \vec c \right|^2 }{|\vec a|^2 \left( |\vec a|^2 |\vec b|^2 - | \vec a \cdot \vec b |^2  \right) } } \sqrt{M^{}_3}
\end{array} \right) \;.
\label{2.3}
\end{eqnarray}

Correspondingly, $M^{}_\nu$ turns out to be given by $M^{}_\nu = U^{}_{\rm L} \Delta M^{-1}_{\rm R} \Delta^T U^{T}_{\rm L}$. It is easy to see that the resulting neutrino mixing matrix can be expressed as $U= U^{}_{\rm L} U^{}_{\rm R}$ with $U^{}_{\rm R}$ being the unitary matrix for diagonalizing $\Delta M^{-1}_{\rm R} \Delta^T$:
\begin{eqnarray}
U^{\dagger}_{\rm R} \Delta M^{-1}_{\rm R} \Delta^T U^{*}_{\rm R} = {\rm diag}(m^{}_1, m^{}_2, m^{}_3) \;.
\label{2.4}
\end{eqnarray}
With the help of this result, one can make the following observation: when $U^{}_{\rm R}$ coincides with $U^{}_{23}$ (or $U^{}_{13}$), which will be the case for $\vec a \cdot \vec b = \vec a \cdot \vec c =0$ (or $\vec a \cdot \vec b = \vec b \cdot \vec c =0$), $U$ will retain the first (or second) column of $U^{}_{\rm L}$ which is in turn proportional to the corresponding column of $M^{}_{\rm D}$. Therefore, in order to obtain a TM1 or TM2 mixing, $M^{}_{\rm D}$ should have such a texture that one column of it is proportional to the first or second column of $U^{}_{\rm TBM}$ while the other two columns are orthogonal to it
\footnote{We point out that such a guiding principle can also be employed to formulate the generic texture of $M^{}_{\rm D}$ that gives a neutrino mixing matrix with one column being of any desired pattern.}.
To be specific, the generic textures of $M^{}_{\rm D}$ that can naturally yield the trimaximal mixings can be parameterized as
\begin{eqnarray}
&& {\rm TM1}: \hspace{1cm} M^{}_{\rm D}= \left( \begin{array}{ccc}
\vspace{0.1cm}
2  l \sqrt{M^{}_1} & m x \sqrt{M^{}_2} & n  y \sqrt{M^{}_3} \cr
-l  \sqrt{M^{}_1} & m (1+x) \sqrt{M^{}_2} & n (1+y) \sqrt{M^{}_3} \cr
l  \sqrt{M^{}_1} & m (1-x)  \sqrt{M^{}_2} & n (1-y) \sqrt{M^{}_3} \cr
\end{array} \right) \; ; \nonumber \\
&& {\rm TM2}: \hspace{1cm} M^{}_{\rm D}= \left( \begin{array}{ccc}
2 l x \sqrt{M^{}_1} & m \sqrt{M^{}_2} & 2 n y \sqrt{M^{}_3} \cr
l(1-x) \sqrt{M^{}_1} & m \sqrt{M^{}_2} & n (1-y) \sqrt{M^{}_3}  \cr
l(1+x) \sqrt{M^{}_1} & - m \sqrt{M^{}_2} & n (1+y)  \sqrt{M^{}_3} \cr
\end{array} \right) \;,
\label{2.5}
\end{eqnarray}
where $l, m, n, x$ and $y$ are generally complex parameters.

Then, we discuss how to realize the textures of $M^{}_{\rm D}$ in Eq.~(\ref{2.5}) by slightly modifying the flavor-symmetry models for realizing the TBM mixing.
Let us first recapitulate the key points of the latter \cite{FS}:
under the specified flavor symmetry (e.g., $A^{}_4$ and $S^{}_4$), three lepton doublets constitute a triplet representation $L=(L^{}_e, L^{}_\mu, L^{}_\tau)$ while three right-handed neutrinos are simply singlets. To break the flavor symmetry in a proper way, three flavon fields $\phi^{}_{J}$ (for $J=1, 2, 3$) are introduced. Each of them is a triplet (with three components $\phi^{}_{J} = [(\phi^{}_{J})^{}_1, (\phi^{}_{J})^{}_2, (\phi^{}_{J})^{}_3]^T$) under the flavor symmetry.
Owing to such a setting, the following dimension-5 operators
\begin{eqnarray}
\sum^{}_{I, J} \frac{ y^{}_{IJ}}{\Lambda}[ \overline L^{}_e (\phi^{}_{J})^{}_1 + \overline L^{}_\mu (\phi^{}_{J})^{}_2 + \overline L^{}_\tau (\phi^{}_{J})^{}_3 ] H N^{}_I \;,
\label{2.6}
\end{eqnarray}
will serve to generate the Dirac neutrino mass terms after the electroweak and flavor symmetries are respectively broken by non-zero vacuum expectation values (VEVs) of the Higgs and flavon fields. Here $y^{}_{IJ}$ are dimensionless coefficients and $\Lambda$ is the energy scale for the flavor-symmetry physics.
In the so-called indirect models \cite{indirect}, the successful realization of the TBM mixing is crucially dependent on the following two practices: three flavon fields are respectively associated with three right-handed neutrinos (i.e., $y^{}_{IJ} = 0$ for $I \neq J$), which can be fulfilled by invoking an auxiliary flavor symmetry; they acquire some particular VEV alignments as
\begin{eqnarray}
\langle \phi^{}_1 \rangle \propto (2, -1, 1)^T \;, \hspace{1cm} \langle \phi^{}_2 \rangle \propto (1, 1, -1)^T \;, \hspace{1cm} \langle \phi^{}_3 \rangle \propto (0, 1, 1)^T \;,
\label{2.7}
\end{eqnarray}
which can be achieved via the so-called F-term alignment mechanism \cite{F-term}. It is then straightforward to verify that $M^{}_{\rm D}$ will take a form as
\begin{eqnarray}
M^{}_{\rm D}= \left( \begin{array}{ccc} \vspace{0.15cm}
2 l \sqrt{M^{}_1} & m \sqrt{M^{}_2}  & 0 \cr \vspace{0.15cm}
- l \sqrt{M^{}_1} & m \sqrt{M^{}_2} & n \sqrt{M^{}_3} \cr
l \sqrt{M^{}_1} & - m \sqrt{M^{}_2} & n \sqrt{M^{}_3} \cr
\end{array} \right)  \;,
\label{2.8}
\end{eqnarray}
which subsequently gives rise to the TBM mixing.

By slightly modifying the above flavor-symmetry models for realizing the TBM mixing, one can
realize the textures of $M^{}_{\rm D}$ in Eq.~(\ref{2.5}): the one that can naturally yield the TM1 mixing can be realized by associating $\langle \phi^{}_1 \rangle$ in Eq.~(\ref{2.7}) with $N^{}_1$ and both $\langle \phi^{}_2 \rangle$ and $\langle \phi^{}_3 \rangle$ with $N^{}_2$ and $N^{}_3$, while the one that can naturally yield the TM2 mixing can be realized by associating $\langle \phi^{}_2 \rangle$ with $N^{}_2$ and both $\langle \phi^{}_1 \rangle$ and $\langle \phi^{}_3 \rangle$ with $N^{}_1$ and $N^{}_3$.
This can be achieved by slightly modifying the charge assignments of the relevant fields under the aforementioned auxiliary flavor symmetry.
Apparently, for the texture of $M^{}_{\rm D}$ that can naturally yield the TM1 (or TM2) mixing, $x$ and $y$ can be viewed as a measure for the relative size between the contributions of $\langle \phi^{}_2 \rangle$ and $\langle \phi^{}_3 \rangle$ (or $\langle \phi^{}_1 \rangle$ and $\langle \phi^{}_3 \rangle$) to the second and third (or first and third) columns of it.

As is known, the S$^{}_4$ group (i.e., the permutation group of four objects) is the unique group that can naturally accommodate the TBM mixing from the group-theoretical consideration \cite{s4}.
Since the trimaximal mixings are some variants of the TBM mixing, the S$^{}_4$ group can also naturally accommodate them \cite{TM2}-\cite{Tanimoto}. The S$^{}_4$ group has 24 elements and 5 irreducible representations: two singlets ${\bf 1}$ and ${\bf 1^\prime}$, one doublet ${\bf 2}$ and two triplets ${\bf 3}$ and ${\bf 3^\prime}$. The matrix forms of the 24 elements in these representations, the Kronecker products of two representations and the Clebsch relations can be found in Ref.~\cite{group} (see also Appendix B of Ref.~\cite{littlest}).

\section{Simplified textures of $M^{}_{\rm D}$ for the TM1 mixing}

In this section, we examine if the parameters of the generic texture of $M^{}_{\rm D}$ that can naturally yield the TM1 mixing can be further reduced, giving more simplified textures of it. The consequences of the phenomenologically-viable simplified textures of $M^{}_{\rm D}$ for the neutrino parameters and leptogenesis will be studied.

\subsection{Phenomenologically-viable simplified textures}

Our analysis will be restricted to the simple but instructive scenario that three elements in the same column of $M^{}_{\rm D}$ share a common phase, which is often the case in the flavor-symmetry models \cite{FS}. Given that an overall rephasing of $M^{}_{\rm D}$ is of no physical meaning, the second column of $M^{}_{\rm D}$ will be taken to be real without loss of generality. Furthermore, considering that the first-column phase ${\rm arg}(l)$ only contributes to $\rho$ additively (see Eqs.~(\ref{3.4}, \ref{3.5})), the first column of $M^{}_{\rm D}$ will also be taken to be real. (In the case of ${\rm arg}(l) \neq 0$, one just needs to make a simple replacement $\rho \to \rho+{\rm arg}(l)$ for our results.) We are therefore left with only one phase parameter, the third-column phase, which will be responsible for both the CP-violating effects at low energies and leptogenesis.
Accordingly, for convenience of the following discussions, the texture of $M^{}_{\rm D}$ that can naturally yield the TM1 mixing is reexpressed as
\begin{eqnarray}
M^{}_{\rm D}= \left( \begin{array}{ccc}
\vspace{0.1cm}
2 l  \sqrt{M^{}_1} & m x \sqrt{M^{}_2} & n e^{{\rm i} \phi} y \sqrt{M^{}_3} \cr
-l  \sqrt{M^{}_1} & m (1+x) \sqrt{M^{}_2} & n e^{{\rm i} \phi} (1+y) \sqrt{M^{}_3} \cr
l  \sqrt{M^{}_1} & m (1-x)  \sqrt{M^{}_2} & n e^{{\rm i} \phi} (1-y) \sqrt{M^{}_3} \cr
\end{array} \right) \; ,
\label{3.1}
\end{eqnarray}
with now $l$, $m$, $n$, $x$ and $y$ being real parameters and $\phi$ the only phase parameter.

Our analysis will be further restricted to the following simplified textures of $M^{}_{\rm D}$, which will be instructive for the model-building exercises: (1) there are some vanishing elements \cite{abelian}; (2) there are some equal elements \cite{FS}. It is easy to see that such textures of $M^{}_{\rm D}$ correspond to some particular values of $x$ and $y$: (1) the value $-1$, 0 or 1 of $x$ ($y$) corresponds to a column pattern as $(-1, 0, 2)^{T}$, $(0, 1, 1)^{T}$ or $(1, 2, 0)^{T}$ which has one vanishing element; (2) the value $-1/2$, 0 or 1/2 of $x$ ($y$) corresponds to a column pattern as $(-1, 1, 3)^{T}$, $(0, 1, 1)^{T}$ or $(1, 3, 1)^{T}$ which have a pair of equal elements. Altogether, the particular values of $x$ and $y$ that are phenomenologically appealing include $-1$, $-1/2$, 0, 1/2 and 1. The column patterns corresponding to them are listed in Table~\ref{Table2}.

The above particular values of $x$ and $y$ are motivated from the simplicity viewpoint. Generally speaking, the simplicity viewpoint is consistent with the symmetry viewpoint: from the top-down viewpoint, a simple symmetry is more likely to lead to a simple texture of the neutrino mass matrix; from the bottom-up viewpoint, a simple texture of the neutrino mass matrix is easier to find a simple symmetry justification.
Of course, a simple texture of the neutrino mass matrix is not always associated with a simple symmetry, because it may arise accidently. For this consideration, we will only consider the particular values of $x$ and $y$ that can find a simple symmetry justification.

In the indirect flavor-symmetry models \cite{indirect}, the particular forms of three columns of $M^{}_{\rm D}$ are identified as the particular VEV alignments of the flavon fields associated with them. The latter are determined by the flavor symmetry and usually preserve some subgroups (i.e., residual symmetries) of it. For the column pattern in Table~\ref{Table2} (which are now identified as the VEV alignments of the flavon fields associated with them), it is found that $(-1, 1, 3)^{T}$, $(0, 1, 1)^{T}$ and $(1, 3, 1)^{T}$ respectively keep invariant under the $e^{}_4$, $d^{}_1$ and $f^{}_1$ elements of the S$^{}_4$ group in the ${\bf 3}$ representation
\begin{eqnarray}
&& e^{}_4 = \left( \begin{array}{ccc}
0 & -1  & 0 \cr
-1 & 0  &  0 \cr
0 & 0 &  1 \cr
\end{array} \right) \;, \hspace{1cm}
d^{}_1 = \left( \begin{array}{ccc}
1 & 0  & 0 \cr
0 & 0  & 1 \cr
0 & 1 &  0 \cr
\end{array} \right) \;,   \hspace{1cm}
f^{}_1 = \left( \begin{array}{ccc}
0 & 0  & 1 \cr
0 & 1  & 0 \cr
1 & 0 &  0 \cr
\end{array} \right) \;,
\label{3.01}
\end{eqnarray}
while $(0, 1, 1)^{T}$ also keeps invariant under the $d^{}_2$ element of the S$^{}_4$ group in the ${\bf 3^\prime}$ representation
\begin{eqnarray}
d^{}_2 = \left( \begin{array}{ccc}
-1 & 0  & 0 \cr
0 & 0  &  1 \cr
0 & 1 &  0 \cr
\end{array} \right) \;.
\label{3.02}
\end{eqnarray}
But $(-1, 0, 2)^{T}$ and $(1, 2, 0)^{T}$ do not keep invariant under any element of the S$^{}_4$ group.
Therefore, we will only consider the column patterns $(-1, 1, 3)^{T}$, $(0, 1, 1)^{T}$ and $(1, 3, 1)^{T}$ (which correspond to the particular values $-1/2$, 0 and $1/2$ of $x$ and $y$), but discard the column patterns $(-1, 0, 2)^{T}$ and $(1, 2, 0)^{T}$ (which correspond to the particular values $-1$ and $1$ of $x$ and $y$).

\begin{table}[t]
\centering
\begin{tabular}{cccccc}
\hline
$x$ ($y$) & $-1$ & $-1/2$ & 0 & 1/2 & 1  \\[1.5pt]
\hline
{\rm pattern} & $(-1, 0, 2)^{T}$ & $(-1, 1, 3)^{T}$ & $(0, 1, 1)^{T}$ & $(1, 3, 1)^{T}$ & $(1, 2, 0)^{T}$  \\ [1.5pt]
\hline
\end{tabular}
\caption{ For the TM1 mixing, the particular values of $x$ ($y$) and the corresponding column patterns. }
\label{Table2}
\end{table}

Before proceeding, let us enumerate the formulas useful for our numerical calculations. For $M^{}_{\rm D}$ in Eq.~(\ref{3.1}), the resulting neutrino mixing matrix can be decomposed as $U^{}_{\rm TM1} = U^{}_{\rm TBM} U^{}_{23}$ (see Eq.~(\ref{5})). Here $U^{}_{23}$ is the unitary matrix for diagonalizing the following matrix
\begin{eqnarray}
M^\prime_\nu & \equiv & U^{\dagger}_{\rm TBM} M^{}_{\rm D} M^{-1}_{\rm R} M^{T}_{\rm D} U^{*}_{\rm TBM} \nonumber \\
& = & \left( \begin{array}{ccc}
\vspace{0.1cm}
6 l^2 & 0 & 0 \cr
\vspace{0.1cm}
0 & 3 m^2 x^2 + 3 n^2 y^2 e^{2{\rm i} \phi} &  \sqrt{6} m^2 x  + \sqrt{6} n^2 y e^{2 {\rm i} \phi} \cr
0 & \sqrt{6} m^2 x  + \sqrt{6} n^2 y e^{2 {\rm i} \phi} &  2 m^2 + 2 n^2 e^{2 {\rm i} \phi} \cr
\end{array} \right) \; .
\label{3.2}
\end{eqnarray}
Its parameters $\theta$ and $\varphi$ can be calculated as
\begin{eqnarray}
\tan{2\theta} = \frac{ 2\left|M^{\prime *}_{22} M^{\prime}_{23} + M^{\prime *}_{23} M^{\prime}_{33} \right| } {\left| M^{\prime}_{33}|^2 - | M^{\prime}_{22}\right|^2 } \;, \hspace{1cm}
\varphi = \arg \left( M^{\prime *}_{22} M^{\prime}_{23} + M^{\prime *}_{23} M^{\prime}_{33} \right) \;,
\label{3.3}
\end{eqnarray}
with $M^\prime_{ij}$ denoting the $ij$ element of $M^\prime_\nu$.
We note that the equality between $x$ and $y$ is denied, which would otherwise lead to the unacceptable $\varphi =0$.
Then, the three mixing angles and $\delta$ can be extracted from $U^{}_{\rm TM1}$ according to the formulas in Eqs.~(\ref{6}, \ref{7}). On the other hand, the resulting neutrino mass eigenvalues are given by
\begin{eqnarray}
&& m^{}_1 e^{ 2 {\rm i} \alpha } = M^{\prime}_{11} = l^2 \;, \nonumber \\
&& m^{}_2 e^{ 2 {\rm i} \beta } = M^{\prime}_{22} \cos^2{\theta} + M^{\prime}_{33} \sin^2{\theta} \hspace{0.05cm} e^{-2{\rm i} \varphi} - M^{\prime}_{23} \sin{2\theta} \hspace{0.05cm} e^{- {\rm i} \varphi} \;, \nonumber \\
&& m^{}_3 e^{ 2 {\rm i} \gamma }  = M^{\prime}_{33} \cos^2{\theta} + M^{\prime}_{22} \sin^2{\theta} \hspace{0.05cm} e^{ 2{\rm i} \varphi} + M^{\prime}_{23} \sin{2\theta} \hspace{0.05cm} e^{ {\rm i} \varphi} \;,
\label{3.4}
\end{eqnarray}
from which $\rho$ and $\sigma$ can be obtained as
\begin{eqnarray}
\rho = \varphi - \delta + \alpha - \gamma  \;, \hspace{1cm} \sigma =  \varphi - \delta + \beta - \gamma \;.
\label{3.5}
\end{eqnarray}
It is easy to see that, under the transformation $\phi \to -\phi$, the results for the CP phases undergo a sign reversal while those for the neutrino masses and mixing angles keep invariant.
Furthermore, $\phi$ has a period of $\pi$ in determining the neutrino parameters (see Eq.~(\ref{3.2})) and the CP asymmetries responsible for leptogenesis (as will be seen from Eq.~(\ref{3.15})).

\begin{figure}
\centering
\includegraphics[width=6.5in]{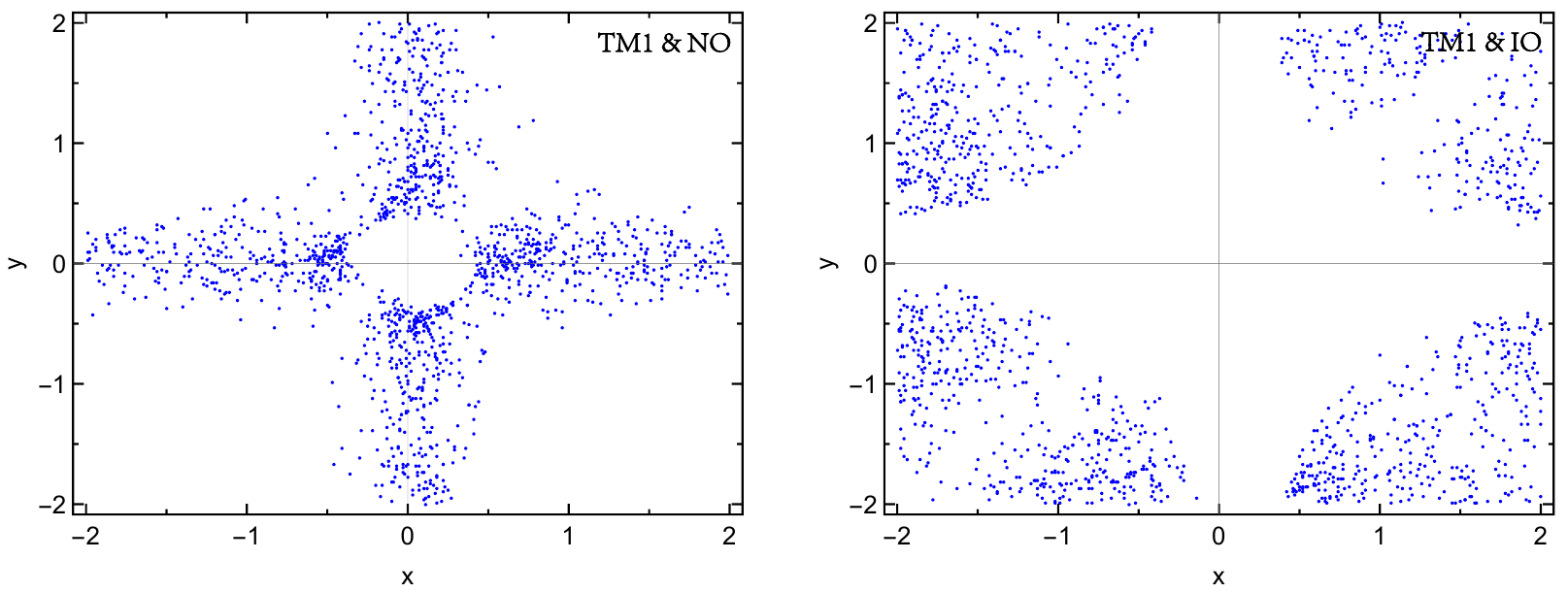}
\caption{ For the TM1 mixing, the values of $y$ versus $x$ that can be phenomenologically viable within the $3\sigma$ level in the NO (left) and IO (right) cases. }
\label{Fig1}
\end{figure}

Now, we confront $M^{}_{\rm D}$ against the experimental results to examine if $x$ and $y$ can take some particular values. Let us first perform the study in the NO case. The left panel of Fig.~\ref{Fig1} shows the values of $y$ versus $x$ that can be phenomenologically viable within the $3\sigma$ level. These results are obtained in a way as follows: for randomly selected values of $x$ and $y$ in the range of $-2$ to 2, $\phi$ in the range of 0 to $\pi$ and $m^{}_1$ in the range of 0.001 eV to 0.1 eV, the values of $l$, $m$ and $n$ are determined by virtue of the following relations for $M^\prime_\nu$ in Eq.~(\ref{3.2})
\begin{eqnarray}
m^{}_1 m^{}_2 m^{}_3 = |{\rm Det}(M^\prime_\nu)| \;, \hspace{1cm} m^2_1+  m^2_2 + m^2_3 = {\rm Tr}( M^{\prime \dagger}_\nu M^\prime_\nu) \;, \hspace{1cm} m^{}_1 = l^2 \;,
\label{3.6}
\end{eqnarray}
where $m^{}_2 = \sqrt{m^2_1 + \Delta m^2_{21}}$ and $m^{}_3 = \sqrt{m^2_1 + \Delta m^2_{31}}$ with $\Delta m^2_{21}$ and $\Delta m^2_{31}$ taking random values in their $3\sigma$ ranges. Then, we check if the resulting values of $\theta$ and $\varphi$ (calculated as in Eq.~(\ref{3.3})) can give some values of $s^2_{13}$ and $s^2_{23}$ (calculated as in Eq.~(\ref{6})) in their $3\sigma$ ranges. (Meanwhile, the values of $\theta^{}_{12}$ and $\delta$ are determined as in Eq.~(\ref{7}).) If yes, then these values of $x$ and $y$ will be recorded. A repetition of the above procedure for enough times yields the results in Fig.~\ref{Fig1}. For the present, we have not taken into account the experimental constraint on $\delta$, but will do so in the following $\chi^2$ calculations.

It is apparent that the results exhibit a symmetry with respect to the interchange $x \leftrightarrow y$. This can be understood as follows: after a successive action of $x \leftrightarrow y$, $m \leftrightarrow n$, $\phi \to -\phi$ and $M^\prime_\nu \to M^\prime_\nu e^{2{\rm i}\phi}$, $M^\prime_\nu$ in Eq.~(\ref{3.2}) keeps invariant except for the replacement $l^2 \to l^2 e^{2{\rm i} \phi}$. This means that the results of $(x, y) = (y^{}_0, x^{}_0)$ can be obtained from those of $(x, y) = (x^{}_0, y^{}_0)$ by making the replacements $\phi \to -\phi$ and $\rho \to \rho + \phi$, where $x^{}_0$ and $y^{}_0$ are any given values of $x$ and $y$. For this reason, we will just consider the $x<y$ cases.
Furthermore, there is a connection between the results of $(x, y) = (x^{}_0, y^{}_0)$ and those of $(x, y) = (-y^{}_0, -x^{}_0)$: after a successive action of $x \leftrightarrow - y$, $m \leftrightarrow n$ and $M^\prime_\nu \to M^\prime_\nu e^{-2{\rm i}\phi}$, $M^\prime_\nu$ in Eq.~(\ref{3.2}) becomes
\begin{eqnarray}
M^\prime_\nu = \left( \begin{array}{ccc}
\vspace{0.1cm}
6 l^2 e^{-2{\rm i}\phi} & 0 & 0 \cr
\vspace{0.1cm}
0 & 3 m^2 x^2 + 3 n^2 y^2 e^{-2{\rm i} \phi} &  -\left( \sqrt{6} m^2 x  + \sqrt{6} n^2 y e^{-2 {\rm i} \phi} \right) \cr
0 & -\left( \sqrt{6} m^2 x  + \sqrt{6} n^2 y e^{-2 {\rm i} \phi} \right) &  2 m^2 + 2 n^2 e^{-2 {\rm i} \phi} \cr
\end{array} \right) \; .
\label{3.7}
\end{eqnarray}
From Eqs.~(\ref{6}-\ref{7}, \ref{3.3}-\ref{3.5}) it is deduced that the results of $(x, y) = (-y^{}_0, -x^{}_0)$ can be obtained from those of $(x, y) = (x^{}_0, y^{}_0)$ by making the replacements $\rho \to -(\rho + \phi)$, $\sigma \to - \sigma$, $\delta \to \pi - \delta $ and $\Delta s^{2}_{23} \to - \Delta s^{2}_{23}$ (for $\Delta s^{2}_{23} \equiv s^{2}_{23} - 1/2$).

\begin{figure}
\centering
\includegraphics[width=5in]{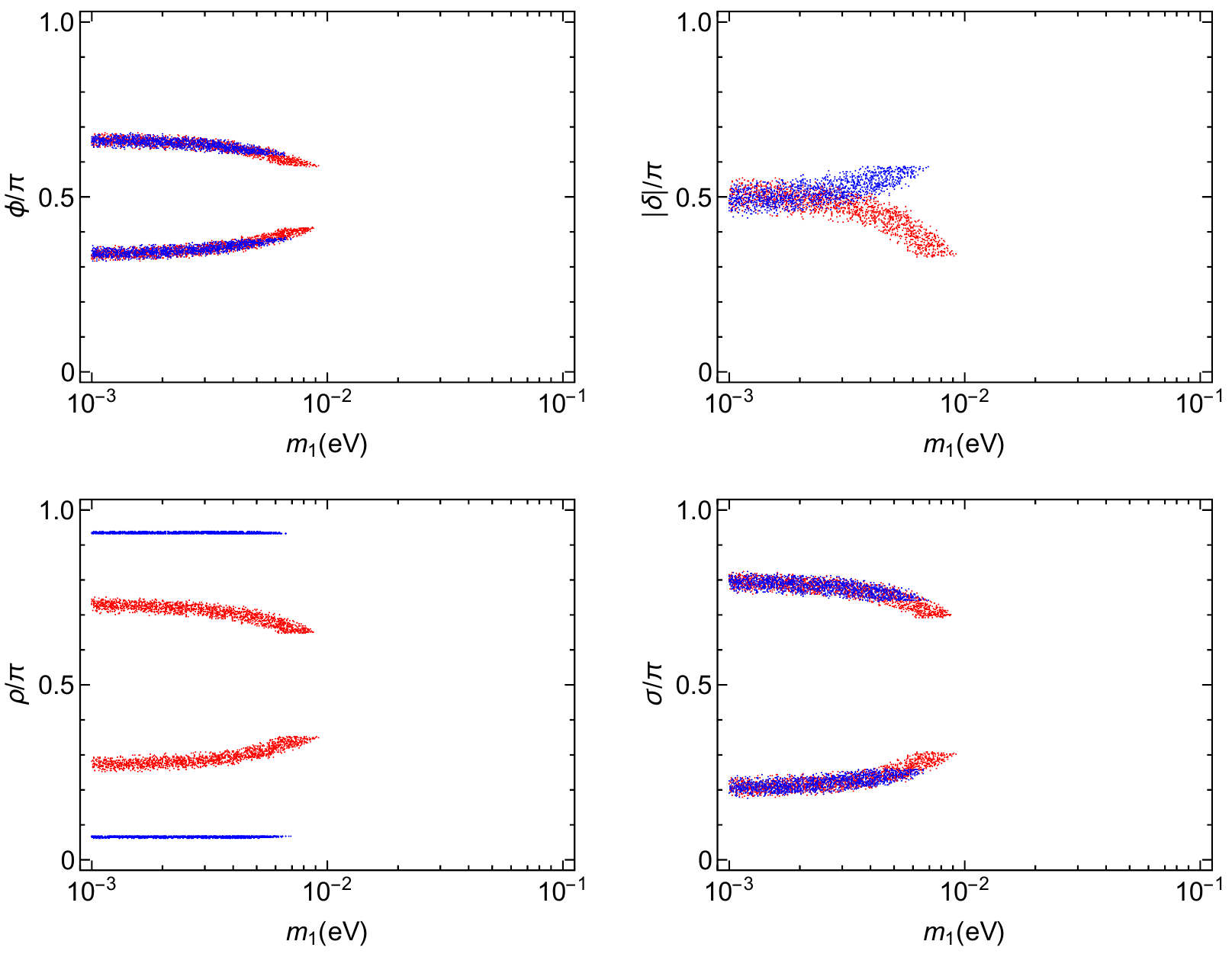}
\caption{ In the NO case with the TM1 mixing, the parameter spaces of $\phi$ versus $m^{}_1$ for $(x, y)= (-1/2, 0)$ (in the red color) and $(0, 1/2)$ (in the blue color) to be phenomenologically viable within the $3\sigma$ level, and the predictions for $|\delta|$, $\rho$ and $\sigma$. }
\label{Fig2}
\end{figure}

Let us consider the possibility that both $x$ and $y$ take some particular values, in which case all the three columns of $M^{}_{\rm D}$ take some simple and instructive patterns. It is found that $(x, y)= (-1/2, 0)$ and $(0, 1/2)$ can be phenomenologically viable within the $3\sigma$ level. Note that the latter case is the  $x \leftrightarrow - y$ counterpart of
the former case, so their results can be related by the replacement rules below Eq.~(\ref{3.7}). For these two cases, Fig.~\ref{Fig2} shows the parameter spaces of $\phi$ and the predictions for $\delta$, $\rho$ and $\sigma$ as functions of $m^{}_1$. These results are tabulated in Table~\ref{Table3}. We see that the predictions for $\delta$ are around $ \pm \pi/2$, in good agreement with the experimental preference for $\delta \sim - \pi/2$. And these two cases can be phenomenologically viable for a negligibly small $m^{}_1$ (realizing an effective minimal seesaw model \cite{minimal}), implying that they can be reduced to the so-called littlest seesaw model \cite{LS, littlest}.

\begin{table}[t]
\centering
\begin{tabular}{ccccccc}
\hline
 $x$ & $y$ & $\phi/\pi$  & $m^{}_1$ (eV) & $\delta/\pi$ & $\rho/\pi$ & $\sigma/\pi$  \\[1.5pt]
\hline
$-1/2$ & 0 & $\pm$(0.32-0.41) & $<$0.009  & $\mp$(0.33-0.56) & $\mp$(0.25-0.35) & $\mp$(0.18-0.31)  \\ [1.5pt]
$0$ & 1/2 & $\pm$(0.32-0.38) & $<$0.007  & $\mp$(0.44-0.59) & $\mp$(0.06-0.07) & $\pm$(0.18-0.26)  \\ [1.5pt]
\hline
\end{tabular}
\caption{ In the NO case with the TM1 mixing, the parameter spaces of $\phi$ for the phenomenologically-viable particular $(x, y)$ combinations, and the predictions for $m^{}_1$, $\delta$, $\rho$ and $\sigma$. }
\label{Table3}
\end{table}

Then, we further examine if $\phi$ can also take a particular value (out of $\pm \pi/6$, $\pm \pi/4$, $\pm \pi/3$), on the basis of particular $(x, y)$ combinations. Table~\ref{Table4} lists the particular $(x, y, \phi)$ combinations that can be phenomenologically viable within the $3\sigma$ level and their predictions for the neutrino parameters at $\chi^2_{\rm min}$.
In the literature, the $\chi^2$ value for a set of neutrino-parameter values is usually calculated as
\begin{eqnarray}
\chi^2 = \sum_i \left( \frac{ {\mathcal O}^{}_{i} - \overline {\mathcal O}^{}_{i} }{ \sigma^{}_{i } } \right)^2 \; ,
\label{3.8}
\end{eqnarray}
where the sum is over three mixing angles, two neutrino mass squared differences and $\delta$, and ${\mathcal O}^{}_i$, $\overline {\mathcal O}^{}_i$ and $\sigma^{}_{i }$ respectively denote their predicted values, best-fit values and $1\sigma$ errors. But this formula has not taken account of the correlations among different parameters. For completeness, in obtaining the results in Table~\ref{Table4} we have made use of the $\chi^2$ tables given at \cite{chi2} which contain such information.

\begin{table}[t]
\centering
\footnotesize
\begin{tabular}{cccccccccccccc}
\hline
$x$ & $y$ & $\phi/\pi$ & $\chi^2_{\rm min}$ & $m^{}_1$ & $\Delta m^2_{21}$ & $|\Delta m^2_{31}|$ & $s^2_{12}$ & $s^2_{13}$ & $s^2_{23}$ & $\delta/\pi$ & $\rho/\pi$ & $\sigma/\pi$ & $|(M^{}_\nu)^{}_{ee}|$ \\[1.5pt]
\hline
$-1/2$ & 0 & 1/3 & 4.0 & 0.001 & 7.44 & 2.52 & 0.318 & 0.02283 & 0.486 & 1.48 & 0.73 & 0.80 & 3.4 \\ [1.5pt]
$0$ & 1/2 & 1/3 & 2.3 & 0.001 & 7.41 & 2.51 & 0.318 & 0.02279 & 0.514 & 1.52 & 0.93 & 0.20 & 2.4 \\ [1.5pt]
\hline
\end{tabular}
\caption{ In the NO case with the TM1 mixing, the predictions of the phenomenologically-viable particular $(x, y, \phi)$ combinations for the neutrino parameters at $\chi^2_{\rm min}$.
The units of $m^{}_1$, $\Delta m^2_{21}$, $|\Delta m^2_{31}|$ and $|(M^{}_\nu)^{}_{ee}|$ are eV, $10^{-5}$ eV$^2$, $10^{-3}$ eV$^2$ and $10^{-3}$ eV, respectively. }
\label{Table4}
\end{table}

As for the IO case, the right panel of Fig.~\ref{Fig1} shows the values of $y$ versus $x$ that can be phenomenologically viable within the $3\sigma$ level. It turns out that $x$ and $y$ are not allowed to take the considered particular values simultaneously.

\subsection{Consequences for leptogenesis}

Let us proceed to study the consequences of the above phenomenologically-viable particular $(x, y, \phi)$ combinations for leptogenesis. As is known, the seesaw model via the leptogenesis mechanism offers an appealing explanation for the baryon asymmetry of the Universe \cite{planck}
\begin{eqnarray}
Y^{}_{\rm B} \equiv \frac{n^{}_{\rm B}-n^{}_{\rm \bar B}}{s} = (8.67 \pm 0.15) \times 10^{-11}  \;,
\label{3.9}
\end{eqnarray}
with $n^{}_{\rm B}$ ($n^{}_{\rm \bar B}$) being the baryon (anti-baryon) number density and $s$ the entropy density. This mechanism proceeds in a way as follows \cite{yanagida,review}: a lepton asymmetry $Y^{}_{\rm L} \equiv (n^{}_{\rm L}-n^{}_{\rm \bar L})/s$ is firstly generated during the decays of the right-handed neutrinos{\footnote{Note that in the flavor-symmetry models the decays of the flavons may also contribute to the generation of the baryon asymmetry \cite{flavon}. This is beyond the scope of the present article. Anyway, the energy scale where the flavor-symmetry physics (i.e., the flavons) resides can be much higher than the right-handed neutrino mass scale where leptogenesis takes place. In this case, the flavons will be decoupled from leptogenesis.}}, and then partially converted into the baryon asymmetry through the sphaleron process \cite{sphaleron}. According to the temperature where leptogenesis takes place (approximately the mass of the right-handed neutrino responsible for leptogenesis), there are three distinct leptogenesis regimes \cite{flavor}. (1) Unflavored regime: in the temperature range above $10^{12}$ GeV, the charged-lepton Yukawa $y^{}_\alpha$ interactions have not yet entered thermal equilibrium, so three lepton flavors are indistinguishable and thus should be treated in a universal way. (2) Two-flavor regime: in the temperature range $10^{9}$---$10^{12}$ GeV, the $y^{}_\tau$-related interactions are in thermal equilibrium, making the $\tau$ flavor distinguishable from the other two flavors which remain indistinguishable. In this regime, the $\tau$ flavor and a superposition of the $e$ and $\mu$ flavors should be treated separately. (3) Three-flavor regime: in the temperature range below $10^{9}$ GeV where the $y^{}_\mu$-related interactions are also in thermal equilibrium, all the three flavors are distinguishable and should be treated separately.
It is well known that the requirement for leptogenesis to be viable places a lower bound $\sim 10^{9}$ GeV for the right-handed neutrino masses \cite{DI}, unless they are nearly degenerate \cite{resonant} --- a possibility beyond the scope of the current paper. Hence we just need to consider the unflavored and two-flavor regimes in the following discussions.

Generally speaking, the final baryon asymmetry is mainly owing to the lightest right-handed neutrino, because the lepton asymmetries generated in the decays of heavier right-handed neutrinos are prone to be erased by the lepton-number-violating interactions of lighter right-handed neutrinos. In the unflavored regime, the baryon asymmetry contributed by $N^{}_I$ is given by
\begin{eqnarray}
Y^{}_{I\rm B} = c r \varepsilon^{}_I \kappa(\widetilde m^{}_I)  \;,
\label{3.10}
\end{eqnarray}
where $c = -28/79$ quantifies the conversion efficiency from the lepton asymmetry to the baryon asymmetry through the sphaleron process \cite{relation}, and $r \simeq 3.9 \times 10^{-3}$ is the ratio of the equilibrium number density of $N^{}_I$ to the entropy density at the temperature above $M^{}_I$. $\varepsilon^{}_I$ is the total CP asymmetry for the decays of $N^{}_I$
\begin{eqnarray}
\varepsilon^{}_{I}
& = & \frac{1}{8\pi (M^\dagger_{\rm D}
M^{}_{\rm D})^{}_{II} v^2} \sum^{}_{J \neq I} {\rm Im}\left[ (M^\dagger_{\rm D} M^{}_{\rm D})^{2}_{IJ}\right] {\cal F} \left( \frac{M^2_J}{M^2_I} \right) \;,
\label{3.11}
\end{eqnarray}
which is a sum of the flavored CP asymmetries \cite{yanagida,cp}
\begin{eqnarray}
\varepsilon^{}_{I \alpha} & \equiv & \frac{ \left[ \Gamma(N^{}_I \to L^{}_\alpha + H) - \Gamma(N^{}_I \to \overline{L^{}_\alpha} + \overline{H}) \right]}{ \sum^{}_\alpha \left[\Gamma(N^{}_I \to L^{}_\alpha + H) + \Gamma(N^{}_I \to \overline{L^{}_\alpha} + \overline{H} ) \right] }  \nonumber \\
& = & \frac{1}{8\pi (M^\dagger_{\rm D}
M^{}_{\rm D})^{}_{II} v^2} \sum^{}_{J \neq I} \left\{ {\rm Im}\left[(M^*_{\rm D})^{}_{\alpha I} (M^{}_{\rm D})^{}_{\alpha J}
(M^\dagger_{\rm D} M^{}_{\rm D})^{}_{IJ}\right] {\cal F} \left( \frac{M^2_J}{M^2_I} \right) \right. \nonumber \\
&  &
+ \left. {\rm Im}\left[(M^*_{\rm D})^{}_{\alpha I} (M^{}_{\rm D})^{}_{\alpha J} (M^\dagger_{\rm D} M^{}_{\rm D})^*_{IJ}\right] {\cal G}  \left( \frac{M^2_J}{M^2_I} \right) \right\} \; ,
\label{3.12}
\end{eqnarray}
with $v =174$ GeV being the Higgs vacuum expectation value, ${\cal F}(x) = \sqrt{x} \{(2-x)/(1-x)+ (1+x) \ln [x/(1+x)] \}$ and ${\cal G}(x) = 1/(1-x)$. Finally, $\kappa(\widetilde m^{}_I)$ is the efficiency factor accounting for the washout effects due to the inverse-decay and lepton-number-violating scattering processes \cite{giudice}. Its value is determined by the washout mass parameter $\widetilde m^{}_I$, which is a sum of the flavored washout mass parameters
\begin{eqnarray}
\widetilde m^{}_{I \alpha} = \frac{|(M^{}_{\rm D})^{}_{\alpha I}|^2}{M^{}_I} \; .
\label{3.13}
\end{eqnarray}
In the two-flavor regime, the baryon asymmetry receives two contributions from $\varepsilon^{}_{I \tau}$ and $\varepsilon^{}_{I \gamma} = \varepsilon^{}_{I e} + \varepsilon^{}_{I \mu}$ which are subject to different washout effects controlled by $\widetilde m^{}_{I \tau}$ and $\widetilde m^{}_{I \gamma} = \widetilde m^{}_{I e} + \widetilde m^{}_{I \mu}$ \cite{flavor}
\begin{eqnarray}
Y^{}_{I\rm B}
= c r \left[ \varepsilon^{}_{I \tau} \kappa \left(\frac{390}{589} \widetilde m^{}_{I \tau} \right) + \varepsilon^{}_{I \gamma} \kappa \left(\frac{417}{589} \widetilde m^{}_{I \gamma} \right) \right] \;.
\label{3.14}
\end{eqnarray}

\begin{table}[t]
\centering
\footnotesize
\begin{tabular}{ccccccccccccc}
\hline
$x$ & $y$ & $\phi/\pi$ & {\bf (1a)/(2a) } & {\bf (1b)}  & {\bf (2b)} & {\bf (2c)} & {\bf (1a$^\prime$)/(2a$^\prime$)} & {\bf (1b$^\prime$)}  & {\bf (2b$^\prime$)} & {\bf (2c$^\prime$) }  \\[1.5pt]
\hline
$-1/2$ & 0 & 1/3 & $-$ & 0.23 & $-$ & 0.38 & $-$ & $-$ & $-$ & $-$ \\ [1.5pt]
$0$ & 1/2 & 1/3 & $-$ & 0.94 & $-$ & 1.1 & $-$ & $-$ & $-$ & $-$ \\ [1.5pt]
\hline
\end{tabular}
\caption{ In the NO case with the TM1 mixing, for the phenomenologically-viable particular $(x, y, \phi)$ combinations, the values of $M^{}_2$ in {\bf Scenario (1a)/(2a)}, {\bf (1b)}, {\bf (2b)} and {\bf (2c)} and $M^{}_3$ in {\bf Scenario (1a$^\prime$)/(2a$^\prime$)}, {\bf (1b$^\prime$)}, {\bf (2b$^\prime$)} and {\bf (2c$^\prime$)} for leptogenesis to be viable.
The units of $M^{}_2$ and $M^{}_3$ are $10^{11}$ GeV. }
\label{Table42}
\end{table}

Because of the special form of $M^{}_{\rm D}$ in Eq.~(\ref{3.1}), which leads to $(M^\dagger_{\rm D} M^{}_{\rm D})^{}_{12} = (M^\dagger_{\rm D} M^{}_{\rm D})^{}_{13} =0$, the CP asymmetries for the decays of $N^{}_1$ (i.e., $\varepsilon^{}_1$ and $\varepsilon^{}_{1\alpha}$) are vanishing. Hence the final baryon asymmetry can only be owing to $N^{}_2$ or $N^{}_3$, even when $N^{}_1$ is the lightest one \cite{N2}. But it should be noted that the lepton asymmetry generated in the decays of $N^{}_2$ or $N^{}_3$ would be subject to the washout effects from the lepton-number-violating interactions of $N^{}_1$ if it is lighter. Taking account of the interplay between the right-handed neutrino mass spectrum and the flavor effects, there are the following possible scenarios for leptogenesis. For $M^{}_2 < M^{}_1, M^{}_3$, the final baryon asymmetry is mainly owing to $N^{}_2$ and the washout effects from $N^{}_1$ are decoupled. Depending on the comparison between $M^{}_2$ with $10^{12}$ GeV (i.e., the boundary between the unflavored and two-flavor regimes), there are the following two possible scenarios. {\bf Scenario (1a)}: For $M^{}_2 > 10^{12}$ GeV, the final baryon asymmetry $Y^{}_{2 \rm B}$ can be calculated according to Eq.~(\ref{3.10}) with
\begin{eqnarray}
\varepsilon^{}_{2 } =
\frac{M^{}_3 n^2 (2+ 3 xy  )^2 }{8 \pi v^2 ( 2 + 3 x^2 ) } {\cal F} \left( \frac{M^2_3}{M^2_2} \right)  \sin 2\phi \;, \hspace{1cm} \widetilde m^{}_2 = m^2(2+3x^2)  \; .
\label{3.15}
\end{eqnarray}
Apparently, as mentioned in the above, $\phi$ has a period of $\pi$ in determining the CP asymmetries for leptogenesis, and $\phi = \pi/2$ would prohibit a viable leptogenesis.
{\bf Scenario (1b)}: For $M^{}_2 < 10^{12}$ GeV, $Y^{}_{2 \rm B}$ can be calculated according to Eq.~(\ref{3.14}) with
\begin{eqnarray}
\varepsilon^{}_{2 \tau}  =
\frac{M^{}_3 n^2 (1-x) (1-y) ( 2+ 3 xy ) }{8 \pi v^2 (2 + 3 x^2) } {\cal F} \left( \frac{M^2_3}{M^2_2} \right)  \sin 2\phi   \; , \hspace{1cm} \widetilde m^{}_{2\tau} = m^2(1-x)^2 \;,
\label{3.16}
\end{eqnarray}
and $\varepsilon^{}_{2 \gamma}  =  \varepsilon^{}_{2} - \varepsilon^{}_{2 \tau}$ and $\widetilde m^{}_{2\gamma} = \widetilde m^{}_2 - \widetilde m^{}_{2\tau}$.

For $M^{}_1 < M^{}_2 < M^{}_3$, the final baryon asymmetry is also mainly owing to $N^{}_2$, but the washout effects from $N^{}_1$ may become non-negligible. Depending on the comparison between $M^{}_1$ and $M^{}_2$ with $10^{12}$ GeV, there are the following three possible scenarios. {\bf Scenario (2a)}: For $ 10^{12} \ {\rm GeV}<M^{}_1 < M^{}_2$, the washout effects from $N^{}_1$ are along the $\ket{L^{}_1}$ direction in the lepton-flavor space while the lepton asymmetry generated in the decays of $N^{}_2$ is along the $\ket{L^{}_2}$ direction, where $\ket{L^{}_I}$ are the coherent superpositions of $\ket{L^{}_\alpha}$ that couple with $N^{}_I$:
\begin{eqnarray}
\ket{L^{}_I} = \frac{1}{ \sqrt{(M^\dagger_{\rm D} M^{}_{\rm D})^{}_{II}} } \sum^{}_\alpha (M^{}_{\rm D})^*_{\alpha I} \ket{L^{}_\alpha} \;.
\label{3.17}
\end{eqnarray}
Since $\ket{L^{}_1}$ is orthogonal to $\ket{L^{}_2}$ (i.e., $\braket{ L^{}_{1} | L^{}_2} =0$), the washout effects from $N^{}_1$ have no effect on the lepton asymmetry generated in the decays of $N^{}_2$.
Therefore, the results in the present scenario are same as in {\bf Scenario (1a)}.
{\bf Scenario (2b)}: For $M^{}_1 <10^{12} \ {\rm GeV}<M^{}_2$, the washout effects from  $N^{}_1$ are along the $\ket{L^{}_\tau}$ and $\ket{L^{}_{1\gamma}}$ directions with $\ket{L^{}_{I\gamma}}$
being defined as
\begin{eqnarray}
\ket{L^{}_{I \gamma}} = \frac{1}{ \sqrt{\left|(M^{}_{\rm D})^{}_{e I}\right|^2 + \left|(M^{}_{\rm D})^{}_{\mu I}\right|^2} } \left[(M^{}_{\rm D})^*_{e I} \ket{L^{}_e} + (M^{}_{\rm D})^*_{\mu I} \ket{L^{}_\mu}\right] \;,
\label{3.18}
\end{eqnarray}
while the lepton asymmetry generated in the decays of $N^{}_2$ remains to be along the $\ket{L^{}_2}$ direction. Consequently, $Y^{}_{2\rm B}$ can be calculated as
\begin{eqnarray}
Y^{}_{2\rm B} = \left[ p^{}_{2\tau} \exp\left(-\frac{3\pi \widetilde m^{}_{1\tau}}{8 m^{}_*}\right) + p^{}_{21\gamma} \exp\left(-\frac{3\pi \widetilde m^{}_{1\gamma}}{8 m^{}_*}\right) + 1- p^{}_{2\tau} - p^{}_{21\gamma} \right] \left[c r \varepsilon^{}_2 \kappa(\widetilde m^{}_2) \right] \;,
\label{3.19}
\end{eqnarray}
with $m^{}_* \simeq 1.1 \times 10^{-3}$ eV and
\begin{eqnarray}
&& p^{}_{2\tau} = \left|\braket{ L^{}_{\tau} | L^{}_2}\right|^2 = \frac{\left|(M^{}_{\rm D})^{}_{\tau 2}\right|^2}{ (M^\dagger_{\rm D} M^{}_{\rm D})^{}_{22} }  \;,
\nonumber \\
&& p^{}_{2 1\gamma} = \left|\braket{ L^{}_{1\gamma} | L^{}_2}\right|^2 = \frac{\left|(M^{}_{\rm D})^{}_{e 1} (M^{}_{\rm D})^{*}_{e 2} + (M^{}_{\rm D})^{}_{\mu 1} (M^{}_{\rm D})^{*}_{\mu 2} \right|^2 }{ (M^\dagger_{\rm D} M^{}_{\rm D} )^{}_{22} \left[\left|(M^{}_{\rm D})^{}_{e 1}\right|^2 + \left|(M^{}_{\rm D})^{}_{\mu 1}\right|^2\right]} \;.
\label{3.20}
\end{eqnarray}
For the form of $M^{}_{\rm D}$ in Eq.~(\ref{3.1}), one arrives at
\begin{eqnarray}
p^{}_{2\tau} = \frac{(1-x)^2}{2+3x^2} \;, \hspace{1cm}
p^{}_{2 1\gamma} = \frac{(1-x)^2}{5(2+3x^2) }  \;, \hspace{1cm} \widetilde m^{}_{1\tau} = \frac{1}{6} m^{}_1 \;, \hspace{1cm} \widetilde m^{}_{1\gamma} = \frac{5}{6} m^{}_1 \;.
\label{3.21}
\end{eqnarray}
{\bf Scenario (2c)}: For $M^{}_1 < M^{}_2 < 10^{12}$ GeV, the washout effects from  $N^{}_1$ are along the $\ket{L^{}_\tau}$ and $\ket{L^{}_{1\gamma}}$ directions
while the lepton asymmetries generated in the decays of $N^{}_2$ are along the $\ket{L^{}_\tau}$ and $\ket{L^{}_{2\gamma}}$ directions. Accordingly, $Y^{}_{2\rm B}$ can be calculated as
\begin{eqnarray}
Y^{}_{2\rm B}  =  c r \left\{ \varepsilon^{}_{2\tau}  \kappa \left(\frac{390}{589} \widetilde m^{}_{2 \tau} \right) \exp\left(-\frac{3\pi \widetilde m^{}_{1\tau}}{8 m^{}_*}\right)  + \varepsilon^{}_{2\gamma} \kappa \left(\frac{417}{589} \widetilde m^{}_{2 \gamma} \right) \left[ \left(1-p^{}_{2\gamma 1\gamma}\right)
+ p^{}_{2\gamma1\gamma} \exp\left(-\frac{3\pi \widetilde m^{}_{1\gamma}}{8 m^{}_*}\right) \right] \right\} \;,
\label{3.22}
\end{eqnarray}
with
\begin{eqnarray}
p^{}_{2\gamma 1\gamma} \equiv \left|\braket{ L^{}_{1\gamma} | L^{}_{2\gamma}}\right|^2 = \frac{  \left| (M^{}_{\rm D})^{}_{e 1} (M^{}_{\rm D})^{*}_{e 2} + (M^{}_{\rm D})^{}_{\mu 1} (M^{}_{\rm D})^{*}_{\mu 2}\right|^2 }{ \left[\left|(M^{}_{\rm D})^{}_{e 1}\right|^2 + \left|(M^{}_{\rm D})^{}_{\mu 1}\right|^2 \right] \left[\left|(M^{}_{\rm D})^{}_{e 2}\right|^2 + \left|(M^{}_{\rm D})^{}_{\mu 2}\right|^2 \right]} \;.
\label{3.23}
\end{eqnarray}
For the form of $M^{}_{\rm D}$ in Eq.~(\ref{3.1}), one arrives at
\begin{eqnarray}
p^{}_{2\gamma 1\gamma} = \frac{(1-x)^2}{5(1+ 2x+2x^2) } \;.
\label{3.24}
\end{eqnarray}
For {\bf Scenario (2b)} and {\bf (2c)}, if $m^{}_1$ is so small that $\widetilde m^{}_{1\tau}$ and $\widetilde m^{}_{1\gamma}$ are much smaller than $m^{}_*$, then the washout effects from $N^{}_1$  would be very weak. Even if $\widetilde m^{}_{1\tau}$ and $\widetilde m^{}_{1\gamma}$ are much larger than $m^{}_*$, a considerable part of the lepton asymmetry generated in the decays of $N^{}_2$ can survive the washout effects from $N^{}_1$ provided that $1 - p^{}_{2\tau} - p^{}_{21\gamma}$ and $1-p^{}_{2\gamma 1\gamma}$ are not too small (e.g., one has $1 - p^{}_{2\tau} - p^{}_{21\gamma} \sim 1$ and $1-p^{}_{2\gamma 1\gamma}$ for $x \sim 1$).

There are also some scenarios where the roles of $N^{}_2$ and $N^{}_3$ are interchanged, which are correspondingly labelled as {\bf (1a$^\prime$)}, {\bf (1b$^\prime$)}, {\bf (2a$^\prime$)}, {\bf (2b$^\prime$)} and {\bf (2c$^\prime$)}. For example, in {\bf Scenario (1a$^\prime$)} one has $10^{12} \ {\rm GeV} < M^{}_3 < M^{}_1, M^{}_2$. In these scenarios, the final baryon asymmetry can be obtained by making the replacements $2 \to 3$, $x \leftrightarrow y$, $M^{}_2 \leftrightarrow M^{}_3$ and $\phi \to - \phi$ in the above expressions.

For the phenomenologically-viable particular $(x, y, \phi)$ combinations listed in Table~\ref{Table4}, the values of $M^{}_2$ in {\bf Scenario (1a)/(2a)}, {\bf (1b)}, {\bf (2b)} and {\bf (2c)} and $M^{}_3$ in {\bf Scenario (1a$^\prime$)/(2a$^\prime$)}, {\bf (1b$^\prime$)}, {\bf (2b$^\prime$)} and {\bf (2c$^\prime$)} for leptogenesis to be viable are calculated and listed in Table~\ref{Table42}.
The results show that leptogenesis has chance to work successfully only for $M^{}_2 < M^{}_3$. Furthermore, even when $N^{}_1$ is the lightest right-handed neutrino, leptogenesis still has chance to work successfully.

\section{Simplified textures of $M^{}_{\rm D}$ for the TM2 mixing}

In this section, we perform a parallel study for the TM2 mixing. Namely, we examine if the parameters of the generic texture of $M^{}_{\rm D}$ that can naturally yield the TM2 mixing can be further reduced, giving more simplified textures of it, and study the consequences of the phenomenologically-viable simplified textures for the neutrino parameters and leptogenesis. As will be seen, the results for the TM2 mixing have a lot in common with those for the TM1 mixing.

\begin{figure}
\centering
\includegraphics[width=6.5in]{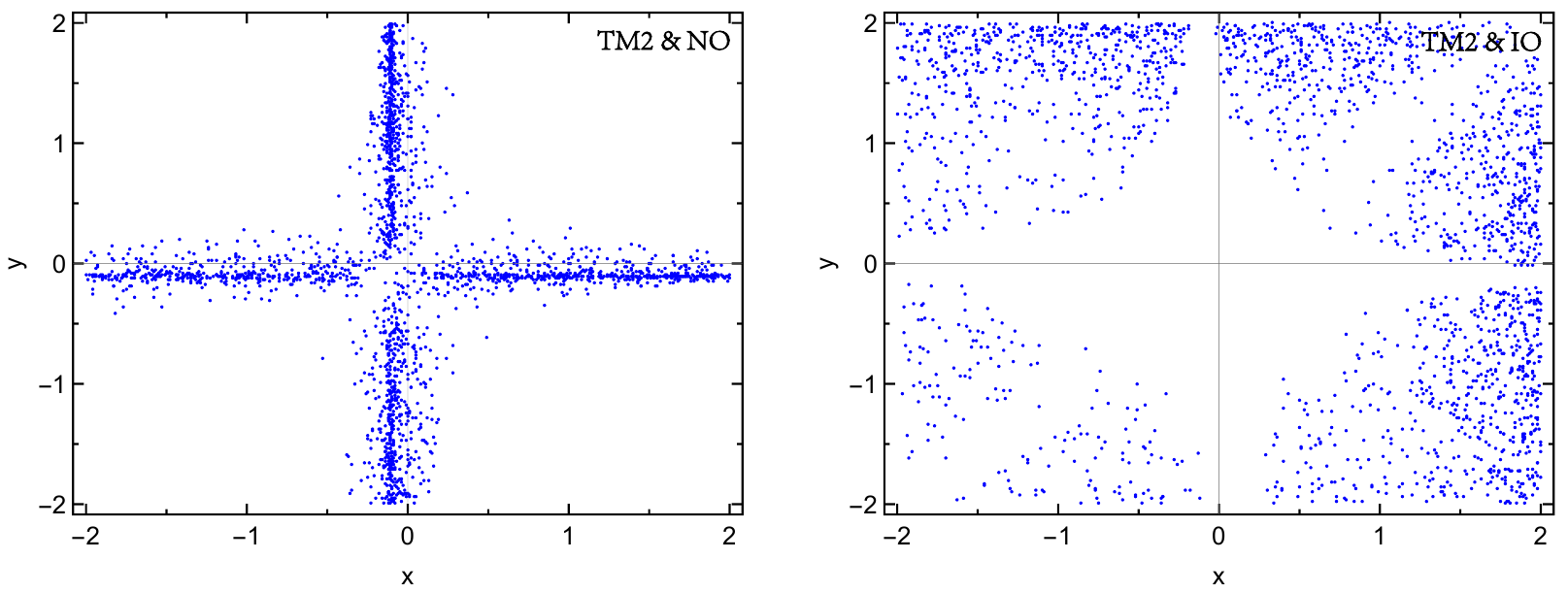}
\caption{ For the TM2 mixing, the values of $y$ versus $x$ that can be phenomenologically viable within the $3\sigma$ level in the NO (left) and IO (right) cases. }
\label{Fig3}
\end{figure}

\subsection{Phenomenologically-viable simplified textures}

For a consideration similar to that given at the beginning of section 3, the first and second columns of $M^{}_{\rm D}$ are taken to be real while the third-column elements share a common phase. Accordingly,  for convenience of the following discussions, the texture of $M^{}_{\rm D}$ that can naturally yield the TM2 mixing is reexpressed as
\begin{eqnarray}
M^{}_{\rm D}= \left( \begin{array}{ccc}
2 l x \sqrt{M^{}_1} & m \sqrt{M^{}_2} & 2 n e^{{\rm i} \phi} y \sqrt{M^{}_3} \cr
l(1-x) \sqrt{M^{}_1} & m \sqrt{M^{}_2} & n e^{{\rm i} \phi} (1-y) \sqrt{M^{}_3}  \cr
l(1+x) \sqrt{M^{}_1} & - m \sqrt{M^{}_2} & n e^{{\rm i} \phi} (1+y)  \sqrt{M^{}_3} \cr
\end{array} \right) \;,
\label{4.1}
\end{eqnarray}
with now $l$, $m$, $n$, $x$ and $y$ being real parameters and $\phi$ the only phase parameter.
From the simplicity viewpoint, the particular values of $x$ and $y$ that are phenomenologically appealing include $-1$, $-1/2$, 0, 1/2 and 1. The column patterns corresponding to them are listed in Table~\ref{Table5}.
From the symmetry viewpoint, it is found that $(-1, 1, 0)^{T}$, $(-1, 2, 1)^{T}$, $(0, 1, 1)^{T}$, $(1, 1, 2)^{T}$ and  $(1, 0, 1)^{T}$ respectively keep invariant under the $e^{}_4$, $f^{}_3$, $d^{}_1$, $e^{}_1$ and $f^{}_1$ elements of the S$^{}_4$ group in the ${\bf 3}$ representation (see Eq.~(\ref{3.01}) for the matrix forms of $e^{}_4$, $d^{}_1$ and $f^{}_1$)
\begin{eqnarray}
f^{}_3 = \left( \begin{array}{ccc}
0 & 0  & -1 \cr
0 & 1  &  0 \cr
-1 & 0 &  0 \cr
\end{array} \right) \;, \hspace{1cm}
e^{}_1 = \left( \begin{array}{ccc}
0 & 1  & 0 \cr
1 & 0  & 0 \cr
0 & 0 &  1 \cr
\end{array} \right) \;,
\label{5.1}
\end{eqnarray}
while $(-1, 1, 0)^{T}$, $(0, 1, 1)^{T}$, and  $(1, 0, 1)^{T}$ respectively keep invariant under the $e^{}_1$, $d^{}_2$ and $f^{}_3$ elements of the S$^{}_4$ group in the ${\bf 3^\prime}$ representation
\begin{eqnarray}
e^{}_1 = \left( \begin{array}{ccc}
0 & -1  & 0 \cr
-1 & 0  &  0 \cr
0 & 0 & -1 \cr
\end{array} \right) \;, \hspace{1cm}
d^{}_2 = \left( \begin{array}{ccc}
-1 & 0  & 0 \cr
0 & 0  &  1 \cr
0 & 1 &  0 \cr
\end{array} \right) \;, \hspace{1cm}
f^{}_3 = \left( \begin{array}{ccc}
0 & 0  & 1 \cr
0 & -1  & 0 \cr
1 & 0 &  0 \cr
\end{array} \right) \;.
\label{5.2}
\end{eqnarray}

\begin{table}[t]
\centering
\begin{tabular}{cccccc}
\hline
$x$ ($y$) & $-1$ & $-1/3$ & 0 & 1/3 & 1  \\[1.5pt]
\hline
{\rm pattern} & $(-1, 1, 0)^{T}$ & $(-1, 2, 1)^{T}$ & $(0, 1, 1)^{T}$ & $(1, 1, 2)^{T}$  & $(1, 0, 1)^{T}$   \\ [1.5pt]
\hline
\end{tabular}
\caption{ For the TM2 mixing, the particular values of $x$ ($y$) and the corresponding column patterns. }
\label{Table5}
\end{table}

For $M^{}_{\rm D}$ in Eq.~(\ref{4.1}), the resulting neutrino mixing matrix can be decomposed as $U^{}_{\rm TM2} = U^{}_{\rm TBM} U^{}_{13}$ (see Eq.~(\ref{5})). Here $U^{}_{13}$ is the unitary matrix for diagonalizing
the following matrix
\begin{eqnarray}
M^\prime_\nu = \left( \begin{array}{ccc}
\vspace{0.1cm}
6 l^2 x^2 + 6 n^2 y^2 e^{2{\rm i} \phi} & 0  & 2 \sqrt{3} l^2 x  + 2 \sqrt{3} n^2 y e^{2{\rm i} \phi} \cr
\vspace{0.1cm}
0 & \displaystyle 3 m^2  &  0 \cr
2 \sqrt{3} l^2 x  + 2 \sqrt{3} n^2 y e^{2{\rm i} \phi} & 0 &  2 l^2 + 2 n^2 e^{2 {\rm i} \phi} \cr
\end{array} \right) \;.
\label{4.2}
\end{eqnarray}
Its parameters $\theta$ and $\varphi$ can be calculated as
\begin{eqnarray}
\tan{2\theta} = \frac{ 2\left|M^{\prime *}_{11} M^{\prime}_{13} + M^{\prime *}_{13} M^{\prime}_{33} \right| } {\left| M^{\prime}_{33}|^2 - | M^{\prime}_{11}\right|^2 } \;, \hspace{1cm}
\varphi = \arg \left( M^{\prime *}_{11} M^{\prime}_{13} + M^{\prime *}_{13} M^{\prime}_{33} \right) \;.
\label{4.3}
\end{eqnarray}
Then, the three mixing angles and $\delta$ can be extracted from $U^{}_{\rm TM2}$ according to the formulas in Eqs.~(\ref{6}, \ref{7}). On the other hand, the resulting neutrino mass eigenvalues are given by
\begin{eqnarray}
&& m^{}_1 e^{ 2 {\rm i} \alpha } = M^{\prime}_{11} \cos^2{\theta} + M^{\prime}_{33} \sin^2{\theta} \hspace{0.05cm} e^{-2{\rm i} \varphi} - M^{\prime}_{13} \sin{2\theta} \hspace{0.05cm} e^{- {\rm i} \varphi} \;, \nonumber \\
&& m^{}_2 e^{ 2 {\rm i} \beta } = M^{\prime}_{22} = 3 m^2 \;, \nonumber \\
&& m^{}_3 e^{ 2 {\rm i} \gamma }  = M^{\prime}_{33} \cos^2{\theta} + M^{\prime}_{11} \sin^2{\theta} \hspace{0.05cm} e^{ 2{\rm i} \varphi} + M^{\prime}_{13} \sin{2\theta} \hspace{0.05cm} e^{ {\rm i} \varphi} \;,
\label{4.4}
\end{eqnarray}
from which $\rho$ and $\sigma$ can also be calculated as in Eq.~(\ref{3.5}).

\begin{figure}
\centering
\includegraphics[width=5in]{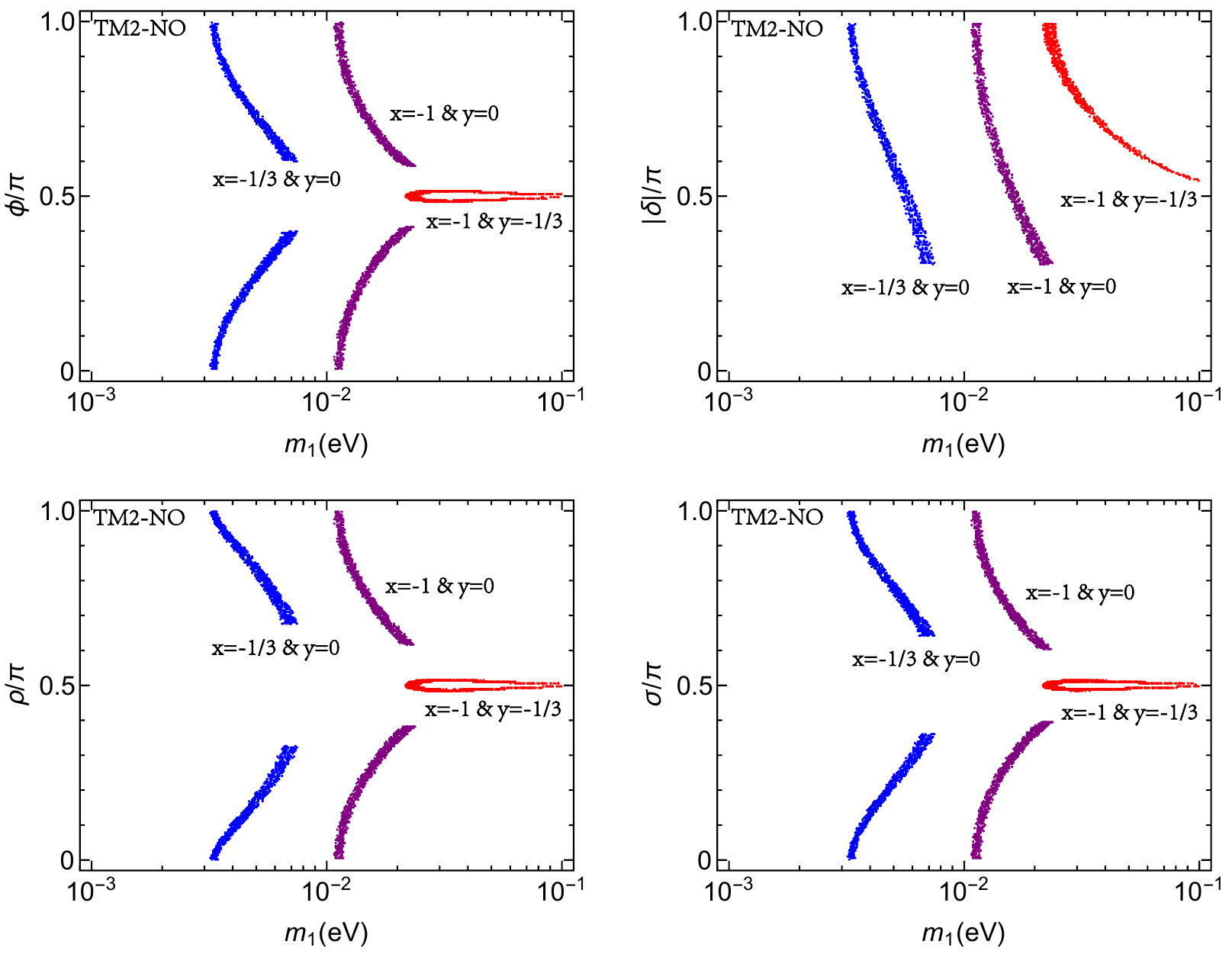}
\caption{ In the NO case with the TM2 mixing, the parameter spaces of $\phi$ versus $m^{}_1$ for $(x, y)= (-1, -1/3)$, $(-1, 0)$ and $(-1/3, 0)$ to be phenomenologically viable within the $3\sigma$ level, and the predictions for $|\delta|$, $\rho$ and $\sigma$.  }
\label{Fig4}
\end{figure}

\begin{figure}
\centering
\includegraphics[width=5in]{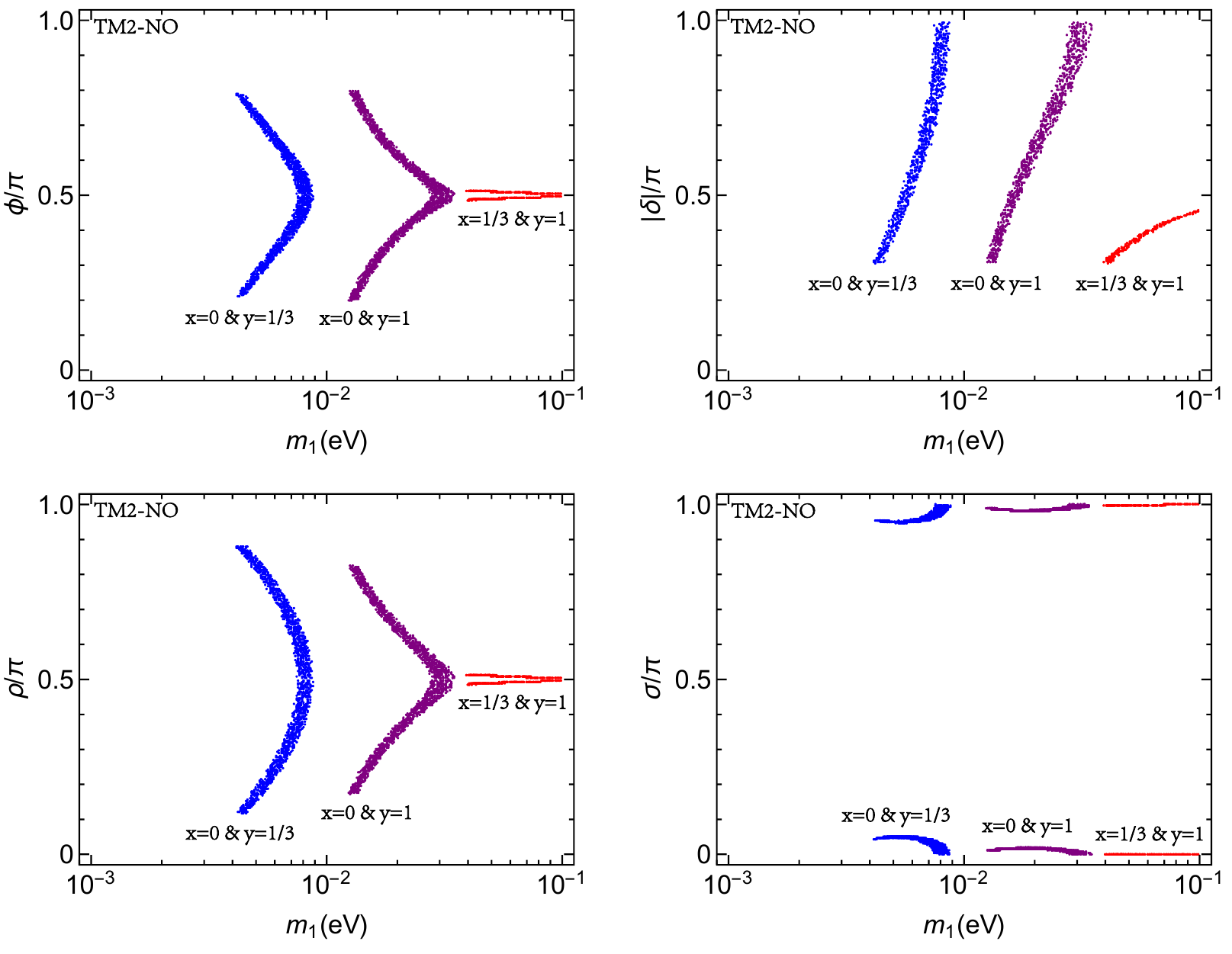}
\caption{ In the NO case with the TM2 mixing, the parameter spaces of $\phi$ versus $m^{}_1$ for $(x, y)= (0, 1/3)$, $(0, 1)$ and $(1/3, 1)$ to be phenomenologically viable within the $3\sigma$ level, and the predictions for $|\delta|$, $\rho$ and $\sigma$.  }
\label{Fig4-2}
\end{figure}

It is analogously deduced that the results of $(x, y)= (y^{}_0, x^{}_0)$ can be obtained from those of $(x, y)= (x^{}_0, y^{}_0)$ by making the replacements $\phi \to -\phi$ and $\sigma \to \sigma + \phi$, so we will just consider the $x<y$ cases. Furthermore, the results of $(x, y)= (- y^{}_0, -x^{}_0)$ can be obtained from those of $(x, y)= (x^{}_0, y^{}_0)$ by making the replacements $\rho \to -\rho$, $\sigma \to - (\sigma + \phi)$, $\delta \to \pi -\delta$ and $\Delta s^{2}_{23} \to - \Delta s^{2}_{23}$.

Now let us consider the possibility that both $x$ and $y$ take some particular values. Fig.~\ref{Fig3} shows the values of $y$ versus $x$ that can be phenomenologically viable within the $3\sigma$ level in the NO (left) and IO (right) cases. The results are obtained in the same way as for the TM1 mixing, except that now the values of $l$, $m$ and $n$ are determined by virtue of the following relations for $M^\prime_\nu$ in Eq.~(\ref{4.2})
\begin{eqnarray}
m^{}_1 m^{}_2 m^{}_3 = |{\rm Det}(M^\prime_\nu)| \;, \hspace{1cm} m^2_1 + m^2_2 + m^2_3 = {\rm Tr}( M^{\prime \dagger}_\nu M^\prime_\nu ) \;, \hspace{1cm} m^{}_2 = 3 m^2 \;.
\label{4.5}
\end{eqnarray}
In the NO case, $(x, y)= (-1, -1/3)$, $(-1, 0)$, $(-1/3, 0)$, $(0, 1/3)$, $(0, 1)$ and $(1/3, 1)$ can be phenomenologically viable within the $3\sigma$ level. Note that the latter three cases are the $x \leftrightarrow -y$ counterparts of the former three cases, so their results can be related by the aforementioned replacement rules. For the former (latter) three cases, Fig.~\ref{Fig4} (Fig.~\ref{Fig4-2})
shows the parameter spaces of $\phi$ and the predictions for $|\delta|$, $\rho$ and $\sigma$ as functions of $m^{}_1$. These results are tabulated in Table~\ref{Table6}. Some remarks are given as follows. (1) The allowed ranges of these parameters are significantly larger than those for the TM1 mixing. This can be attributed to the results below Eq.~(\ref{6}). (2) None of these cases can be phenomenologically viable for a negligibly small $m^{}_1$ and thus accommodated in the minimal seesaw framework. (3) Note that for $(x, y) = (-1, -1/3)$ and $(1/3, 1)$, although $\phi$ is not allowed to exactly take $\pi/2$, we have also listed their results in Table~\ref{Table6}, which might be instructive for the model-building exercises.
It is obvious that $\phi = \pi/2$ would render $M^\prime_\nu$ in Eq.~(\ref{4.2}) real, which in turn leads to trivial $\delta$, $\rho$ and $\sigma$. Nevertheless, for the present two cases, $\phi \simeq \pi/2$ leads to $|\delta| \simeq  \pi/2$. A careful analysis reveals that this is because there occurs a large accidental cancellation for the real part of $M^\prime_{13}$, making its imaginary part (which is controlled by the small deviation of $\phi$ from $\pi/2$ and should have been subdominant) dominant, which subsequently leads to a nearly maximal $|\varphi|$ and thus $|\delta|$.
As for the IO case, only $(x, y)=(-1, -1/3)$ and $(1/3, 1)$ can be phenomenologically viable within the $3\sigma$ level.

\begin{table}[t]
\centering
\begin{tabular}{cccccccc}
\hline
& $x$ & $y$  & $\phi/\pi$ & $m^{}_l$ (eV)  & $\delta/\pi$ & $\rho/\pi$ & $\sigma/\pi$  \\[1.5pt]
\hline
NO & $-$1 & $-$1/3 & $\pm$(0.484-0.499) & $>$0.022  & $\mp$(0.54-0.99) & $\mp$(0.484-0.499) & $\mp$(0.486-0.499)  \\ [1.5pt]
& $-$1 & 0 & $\pm$(0.00-0.41) & 0.011---0.024  & $\mp$(0.31-0.99) & $\mp$(0.00-0.38) & $\mp$(0.00-0.40)  \\ [1.5pt]
& $-$1/3 & 0 & $\pm$(0.01-0.40) & 0.003---0.008  & $\mp$(0.31-0.99) & $\mp$(0.00-0.32) & $\mp$(0.00-0.36)  \\ [1.5pt]
& 0 & 1/3 & $\pm$(0.21-0.50) & 0.004---0.009 & $\mp$(0.31-0.99) & $\pm$(0.12-0.50) & $\mp$(0.00-0.05)  \\ [1.5pt]
& 0 & 1 & $\pm$(0.20-0.50) & 0.013---0.034 & $\mp$(0.31-0.99) & $\pm$(0.17-0.50) & $\mp$(0.00-0.02)  \\ [1.5pt]
& 1/3 & 1 & $\pm$(0.486-0.496) & $>$0.040  & $\mp$(0.31-0.46) & $\pm$(0.486-0.496) & $\pm$(0.000-0.002)  \\ [1.5pt]
IO & $-$1 & $-$1/3 & $\pm$(0.486-0.496) & $>$0.040  & $\pm$(0.31-0.46) & $\mp$(0.486-0.496) & $\mp$(0.487-0.497)  \\ [1.5pt]
& 1/3 & 1 & $\pm$(0.484-0.499) & $>$0.022  & $\pm$(0.54-0.99) & $\pm$(0.484-0.499) & $\pm$(0.000-0.002)  \\ [1.5pt]
\hline
\end{tabular}
\caption{ For the TM2 mixing, the parameter spaces of $\phi$ for the phenomenologically-viable particular $(x, y)$ combinations, and the predictions for $m^{}_l$, $\delta$, $\rho$ and $\sigma$. }
\label{Table6}
\end{table}

Then, we further examine if $\phi$ can also take some particular value, on the basis of particular $(x, y)$ combinations. Table~\ref{Table7} lists the phenomenologically-viable particular $(x, y, \phi)$ combinations and their predictions for the neutrino parameters at $\chi^2_{\rm min}$. Note that for $(x, y) = (-1, -1/3)$ and $(1/3, 1)$, $\phi$ is not allowed to exactly take $\pi/2$ but is very close to it in both the NO and IO cases.

\begin{table}[t]
\centering
\footnotesize
\begin{tabular}{ccccccccccccccc}
\hline
& $x$ & $y$ & $\phi/\pi$ & $\chi^2_{\rm min}$ & $m^{}_1$ & $\Delta m^2_{21}$ & $\Delta m^2_{31}$ & $s^2_{12}$ & $s^2_{13}$ & $s^2_{23}$ & $\delta/\pi$ & $\rho/\pi$ & $\sigma/\pi$ & $|(M^{}_\nu)^{}_{ee}|$ \\[1.5pt]
\hline
NO & $-1$ & $-1/3$ & 0.491 & 3.1 & 0.056 & 7.50 & 2.53 & 0.341 & 0.02163 & 0.539 & 1.38 & 0.51 & 0.51 & 56.5 \\ [1.5pt]
& $-1$ & 0 & 1/6 & 9.6 & 0.012 & 7.50 & 2.55 & 0.341 & 0.02166 & 0.574 & 1.25 & 0.86 & 0.84 & 13.9  \\ [1.5pt]
& $-1$ & 0 & 1/4 & 2.9 & 0.014 & 7.50 & 2.51 & 0.341 & 0.02234 & 0.538 & 1.39 & 0.78 & 0.76 & 15.2 \\ [1.5pt]
& $-1$ & 0 & 1/3 & 5.9 & 0.017 & 7.50 & 2.51 & 0.341 & 0.02247 & 0.490 & 1.53 & 0.70 & 0.68 & 17.6 \\ [1.5pt]
& $-1/3$ & 0 & 1/6 & 11 & 0.004 & 7.50 & 2.56 & 0.341 & 0.02172 & 0.579 & 1.23 & 0.91 & 0.87 & 6.5  \\ [1.5pt]
& $-1/3$ & 0 & 1/4 & 2.9 & 0.005 & 7.50 & 2.52 & 0.341 & 0.02228 & 0.543 & 1.37 & 0.85 & 0.80 & 6.7 \\ [1.5pt]
& $-1/3$ & 0 & 1/3 & 6.1 & 0.006 & 7.50 & 2.51 & 0.341 & 0.02239 & 0.488 & 1.54 & 0.76 & 0.72 & 7.1 \\ [1.5pt]
& 0 & 1/3 & 1/4 & 7.6 & 0.005 & 7.50 & 2.53 & 0.341 & 0.02213 & 0.457 & 1.63 & 0.15 & 0.95 & 4.8  \\ [1.5pt]
& 0 & 1/3 & 1/3 & 4.4 & 0.006 & 7.50 & 2.51 & 0.341 & 0.02260 & 0.512 & 1.46 & 0.24 & 0.95 & 3.6 \\ [1.5pt]
& 0 & 1 & 1/4 & 7.1 & 0.014 & 7.50 & 2.53 & 0.341 & 0.02258 & 0.461 & 1.61 & 0.22 & 0.98 & 11.2 \\ [1.5pt]
& 0 & 1 & 1/3 & 4.3 & 0.017 & 7.50 & 2.51 & 0.341 & 0.02259 & 0.510 & 1.47 & 0.30 & 0.98 & 9.7 \\ [1.5pt]
& 1/3 & 1 & 0.496 & 6.5 & 0.100 & 7.50 & 2.51 & 0.341 & 0.02228 & 0.484 & 1.55 & 0.50 & 0.00 & 32.4 \\ [1.5pt]
IO & $-1$ & $-1/3$ & 0.509 & 9.0 & 0.057 & 7.50 & 2.43 & 0.341 & 0.02302 & 0.463 & 1.61 & 0.49 & 0.49 & 74.9 \\ [1.5pt]
& 1/3 & 1 & 0.515 & 5.9 & 0.033 & 7.50 & 2.43 & 0.341 & 0.02257 & 0.574 & 1.26 & 0.51 & 0.10 & 18.7 \\ [1.5pt]
\hline
\end{tabular}
\caption{ For the TM2 mixing, the predictions of the phenomenologically-viable particular $(x, y, \phi)$ combinations for the neutrino parameters at $\chi^2_{\rm min}$.
The units of $m^{}_l$, $\Delta m^2_{21}$, $|\Delta m^2_{31}|$ and $|(M^{}_\nu)^{}_{ee}|$ are eV, $10^{-5}$ eV$^2$, $10^{-3}$ eV$^2$ and $10^{-3}$ eV, respectively. }
\label{Table7}
\end{table}

\subsection{Consequences for leptogenesis}

\begin{table}[t]
\centering
\footnotesize
\begin{tabular}{cccccccccccc}
\hline
& $x$ & $y$ & $\phi/\pi$ & {\bf (1a)/(2a) } & {\bf (1b)}  & {\bf (2b)} & {\bf (2c)} & {\bf (1a$^\prime$)/(2a$^\prime$)} & {\bf (1b$^\prime$)}  & {\bf (2b$^\prime$)} & {\bf (2c$^\prime$) } \\[1.5pt]
\hline
NO & $-1$ & $-1/3$ & 0.491 & 26 & $-$ & $-$ & $-$ & $-$ & $-$ & $-$ & $-$  \\ [1.5pt]
& $-1$ & 0 & 1/6 & $-$ & 1.1 & $-$ & $-$ & $-$ & $-$ & $-$ & $-$  \\ [1.5pt]
& $-1$ & 0 & 1/4 & $-$ & 0.99 & $-$ & $-$ & $-$ & $-$ & $-$ & $-$  \\ [1.5pt]
& $-1$ & 0 & 1/3 & $-$ & 1.3 & $-$ & $-$ & $-$ & $-$ & $-$ & $-$  \\ [1.5pt]
& $-1/3$ & 0 & 1/6 & $-$ & 0.16 & $-$ & 0.61 & $-$ & $-$ & $-$ & $-$  \\ [1.5pt]
& $-1/3$ & 0 & 1/4 & $-$ & 0.13 & $-$ & 0.48 & $-$ & $-$ & $-$ & $-$  \\ [1.5pt]
& $-1/3$ & 0 & 1/3 & $-$ & 0.13 & $-$ & 0.51 & $-$ & $-$ & $-$ & $-$  \\ [1.5pt]
& 0 & 1/3 & 1/4 & $-$ & 0.94 & 12 & 5.3 & $-$ & $-$ & $-$ & $-$   \\ [1.5pt]
& 0 & 1/3 & 1/3 & $-$ & 1.2 & 17 & 7.1 & $-$ & $-$ & $-$ & $-$ \\ [1.5pt]
& 0 & 1 & 1/4 & $-$ & 3.2 & 46 & $-$ & $-$ & $-$ & $-$ & $-$  \\ [1.5pt]
& 0 & 1 & 1/3 & $-$ & 3.5 & 50 & $-$ & $-$ & $-$ & $-$ & $-$  \\ [1.5pt]
& 1/3 & 1 & 0.496 & 81 & $-$ & $-$ & $-$ & $-$ & $-$ & $-$ & $-$  \\ [1.5pt]
IO & $-1$ & $-1/3$ & 0.509 & $-$ & $-$ & $-$ & $-$ & 27 & $-$ & 36 & $-$  \\ [1.5pt]
& 1/3 & 1 & 0.515 & $-$ & $-$ & $-$ & $-$ & $-$ & 6.8 & 95 & $-$  \\ [1.5pt]
\hline
\end{tabular}
\caption{ In the TM2 mixing scenario, for the phenomenologically-viable particular $(x, y, \phi)$ combinations, the values of $M^{}_1$ in {\bf Scenario (1a)/(2a)}, {\bf (1b)}, {\bf (2b)} and {\bf (2c)} and $M^{}_3$ in {\bf Scenario (1a$^\prime$)/(2a$^\prime$)}, {\bf (1b$^\prime$)}, {\bf (2b$^\prime$)} and {\bf (2c$^\prime$)} for leptogenesis to be viable.
The units of $M^{}_1$ and $M^{}_3$ are $10^{11}$ GeV. }
\label{Table72}
\end{table}

Finally, we study the consequences of the phenomenologically-viable particular $(x, y, \phi)$ combinations for leptogenesis. Because of the special form of $M^{}_{\rm D}$ in Eq.~(\ref{4.1}), which leads to $(M^\dagger_{\rm D} M^{}_{\rm D})^{}_{12} = (M^\dagger_{\rm D} M^{}_{\rm D})^{}_{23} =0$, the CP asymmetries for the decays of $N^{}_2$ are vanishing. Hence the final baryon asymmetry can only be owing to $N^{}_1$ or $N^{}_3$, even when $N^{}_2$ is the lightest one. But the lepton asymmetry generated in the decays of $N^{}_1$ or $N^{}_3$ would be subject to the washout effects from $N^{}_2$ if it is lighter. Taking account of the interplay between the right-handed neutrino mass spectrum and the flavor effects, there are the following possible scenarios for leptogenesis. For $M^{}_1 < M^{}_2, M^{}_3$, the final baryon asymmetry is mainly owing to $N^{}_1$ and the washout effects from $N^{}_2$ are decoupled. Depending on the comparison between $M^{}_1$ with $10^{12}$ GeV, there are the following two possible scenarios. {\bf Scenario (1a)}: For $M^{}_1 > 10^{12}$ GeV, the final baryon asymmetry $Y^{}_{1 \rm B}$ can be calculated according to Eq.~(\ref{3.10}) with
\begin{eqnarray}
\varepsilon^{}_{1}  =
\frac{M^{}_3 n^2 (1+ 3 xy)^2 }{8 \pi v^2 (1 + 3 x^2) } {\cal F} \left( \frac{M^2_3}{M^2_1} \right)  \sin 2\phi   \; ,  \hspace{1cm} \widetilde m^{}_1 = 2 l^2(1+3x^2)  \; .
\label{4.2.1}
\end{eqnarray}
{\bf Scenario (1b)}: For $M^{}_1 < 10^{12}$ GeV, $Y^{}_{1 \rm B}$ can be calculated according to Eq.~(\ref{3.14}) with
\begin{eqnarray}
&& \varepsilon^{}_{1 \tau}  =
\frac{M^{}_3 n^2 (1 +x ) (1+y) (1+ 3 xy) }{8 \pi v^2 (1 + 3 x^2) } {\cal F} \left( \frac{M^2_3}{M^2_1} \right)  \sin 2\phi   \; , \hspace{1cm} \widetilde m^{}_{1\tau} = l^2(1+x)^2 \;,
\label{4.2.2}
\end{eqnarray}
and $\varepsilon^{}_{1 \gamma}  =  \varepsilon^{}_{1} - \varepsilon^{}_{1 \tau}$ and $\widetilde m^{}_{1\gamma} = \widetilde m^{}_1 - \widetilde m^{}_{1\tau}$.

For $M^{}_2 < M^{}_1 < M^{}_3$, the final baryon asymmetry is also mainly owing to $N^{}_1$, but the washout effects from $N^{}_2$ may become non-negligible. Depending on the comparison between $M^{}_1$ and $M^{}_2$ with $10^{12}$ GeV, there are the following three possible scenarios. {\bf Scenario (2a)}: For $ 10^{12} \ {\rm GeV}<M^{}_2 < M^{}_1$, the washout effects from $N^{}_2$ are along the $\ket{L^{}_2}$ direction while the lepton asymmetry generated in the decays of $N^{}_1$ is along the $\ket{L^{}_1}$ direction.
Since $\ket{L^{}_1}$ is orthogonal to $\ket{L^{}_2}$, the washout effects from $N^{}_2$ have no effect on the lepton asymmetry generated in the decays of $N^{}_1$. Therefore, the results in the present scenario are same as in {\bf Scenario (1a)}.
{\bf Scenario (2b)}: For $M^{}_2 <10^{12} \ {\rm GeV} <M^{}_1$, the washout effects from  $N^{}_2$ are along the $\ket{L^{}_\tau}$ and $\ket{L^{}_{2\gamma}}$ directions,
while the lepton asymmetry generated in the decays of $N^{}_1$ remains to be along the $\ket{L^{}_1}$ direction. The final baryon asymmetry can be calculated as in Eqs.~(\ref{3.19}, \ref{3.20}) but with the interchange $1 \leftrightarrow 2$ for the subscripts.
For the form of $M^{}_{\rm D}$ in Eq.~(\ref{4.1}), one arrives at
\begin{eqnarray}
p^{}_{1\tau} = \frac{(1+x)^2}{2(1+3x^2)} \;, \hspace{1cm}
p^{}_{12\gamma} = \frac{(1+x)^2}{4(1+3x^2) }  \;, \hspace{1cm} \widetilde m^{}_{2\tau} = \frac{1}{3} m^{}_2 \;, \hspace{1cm} \widetilde m^{}_{2\gamma} = \frac{2}{3} m^{}_2 \;.
\label{4.2.3}
\end{eqnarray}
{\bf Scenario (2c)}: For $M^{}_2 <M^{}_1 < 10^{12}$ GeV, the washout effects from  $N^{}_2$ are along the $\ket{L^{}_\tau}$ and $\ket{L^{}_{2\gamma}}$ directions
while the lepton asymmetries generated in the decays of $N^{}_1$ are along the $\ket{L^{}_\tau}$ and $\ket{L^{}_{1\gamma}}$ directions. The final baryon asymmetry can be calculated as in Eqs.~(\ref{3.22}, \ref{3.23}) but also with the interchange $1 \leftrightarrow 2$ for the subscripts.
For the form of $M^{}_{\rm D}$ in Eq.~(\ref{4.1}), one has
\begin{eqnarray}
p^{}_{1\gamma 2\gamma} = \frac{(1+x)^2}{2(1-2x+5x^2) } \;.
\label{4.2.4}
\end{eqnarray}
For {\bf Scenario (2b)} and {\bf (2c)}, since $m^{}_2$ is much larger than $m^{}_*$,
the washout effects from $N^{}_2$ are strong. Only when $1 - p^{}_{2\tau} - p^{}_{21\gamma}$ and $1-p^{}_{2\gamma 1\gamma}$ are not too small,
can a considerable part of the lepton asymmetry generated in the decays of $N^{}_1$ survive the washout effects from $N^{}_2$.

There are also some scenarios where the roles of $N^{}_1$ and $N^{}_3$ are interchanged, which are correspondingly labelled as {\bf (1a$^\prime$)}, {\bf (1b$^\prime$)}, {\bf (2a$^\prime$)}, {\bf (2b$^\prime$)} and {\bf (2c$^\prime$)}. In these scenarios, the final baryon asymmetry can be obtained by making the replacements $1 \to 3$, $x \leftrightarrow y$, $M^{}_2 \leftrightarrow M^{}_3$ and $\phi \to - \phi$ in the above expressions.

For the phenomenologically-viable particular $(x, y, \phi)$ combinations listed in Table~\ref{Table7}, the values of $M^{}_1$ in {\bf Scenario (1a)/(2a)}, {\bf (1b)}, {\bf (2b)} and {\bf (2c)} and $M^{}_3$ in {\bf Scenario (1a$^\prime$)/(2a$^\prime$)}, {\bf (1b$^\prime$)}, {\bf (2b$^\prime$)} and {\bf (2c$^\prime$)} for leptogenesis to be viable are calculated and listed in Table~\ref{Table72}.
It is also found that leptogenesis has chance to work successfully only for $M^{}_1 < M^{}_3$ in the NO case but only for $M^{}_3 < M^{}_1$ in the IO case. And leptogenesis still has chance to work successfully even when $N^{}_2$ is the lightest right-handed neutrino.

\section{A concrete flavor-symmetry model}

\begin{table}
\caption{ The transformation properties of the lepton, Higgs and flavon superfields
under the $\rm S^{}_4 \times Z^{}_2 \times Z^{}_3$ symmetries and their R charges. }
\label{Table8}
\vspace{-0.28cm}
$$
\begin{array}{||c||ccccccc||c||cccccccc||}
\hline \hline
 & e^c & \mu^c & \tau^c
& \phi^{}_e & \phi^{}_\mu & \phi^{}_\tau
& H_{d} & L & H^{}_{u} & \phi^{}_{1} & \phi^{}_{2} & \phi^{}_{3} & N^c_{1} & N^c_{2} & N^c_{3}
& \xi  \\  \hline
\hline
\rm  S^{}_4 &  {\bf 1} & {\bf 1} & {\bf 1} & {\bf 3} & {\bf 3} & {\bf 3}  & {\bf 1}
& {\bf 3} & {\bf 1} & {\bf 3'} & {\bf 3} & {\bf 3} & {\bf 1'} & {\bf 1} & {\bf 1} & {\bf 1}
\\ \hline  \hline
\rm  Z^{}_2 & -1 & -1 & -1 & -1 & -1  & -1 & 1 & 1 & 1 & 1 & 1 & 1 & 1 & 1
& 1 & 1  \\\hline
\rm Z^{}_3 & \omega  & \omega^2 & 1 & \omega^2 & \omega & 1 & 1
& 1 & 1 & 1 & 1 & \omega^2 & 1  & 1 & \omega & \omega \\ \hline
{\rm R} &  1 & 1 & 1 & 0 & 0 & 0 & 0 & 1 & 0 & 0 & 0 & 0 & 1 & 1 & 1 & 0  \\
\hline
\end{array}
$$
\end{table}

In this section, we give a concrete S$^{}_4$-flavor-symmetry model that can realize one representation of the simplified textures of $M^{}_{\rm D}$ obtained in the above: the simplified texture of $M^{}_{\rm D}$ in Eq.~(\ref{4.1}) obtained by taking $x=1$ and $y=0$
\begin{eqnarray}
M^{}_{\rm D}= \left( \begin{array}{ccc}
l & m & 0 \cr
0 & m & n e^{{\rm i} \phi}  \cr
l & - m & n e^{{\rm i} \phi}  \cr
\end{array} \right) \;,
\label{5.3}
\end{eqnarray}
while the other ones can be realized analogously.
The model employs $\rm S^{}_4 \times Z^{}_2 \times Z^{}_3$ as the flavor symmetries. And Table~\ref{Table8} gives the transformation properties of the related fields under them. Here the auxiliary Z$^{}_2$ symmetry is used to distinguish the flavon fields associated with the charged-lepton and neutrino sectors.
And the auxiliary $\rm Z^{}_3$ symmetry is introduced to further distinguish the flavon fields associated with different flavors in the same sector. Furthermore, as will be seen soon, it can also help us achieve $\phi = -\pi/3$. Finally, in order to justify the flavon VEV alignments in Eq.~(\ref{5.5}) by means of the F-term alignment mechanism \cite{F-term}, which invokes the R symmetry (by which the superpotential terms are required to carry an R charge of 2) of supersymmetric theories, the model is embedded in the supersymmetry (SUSY) framework.
Under the above setup, the superpotential terms relevant for the lepton masses are given by
\begin{eqnarray}
W & = & \frac{y^{}_1}{\Lambda} H^{}_u \left(L \cdot \phi^{}_1\right) N^c_{1}
+ \frac{y^{}_2}{\Lambda} H^{}_u \left(L \cdot \phi^{}_2\right) N^c_{2}
+ \frac{y^{}_3}{\Lambda} H^{}_u \left(L \cdot \phi^{}_2\right) N^c_{2} \nonumber \\
&& + M^{}_1 N^c_{1} N^c_{1} + M^{}_2 N^c_{2} N^c_{2} + \xi N^c_{3} N^c_{3} \nonumber \\
&& +\frac{y^{}_e}{\Lambda}  H^{}_d \left( L \cdot \phi^{}_e\right) e^c
+ \frac{y^{}_\mu}{\Lambda}  H^{}_d \left(L \cdot \phi^{}_{\mu}\right) \mu^c
+\frac{y^{}_\tau}{\Lambda} H^{}_d \left(L \cdot \phi^{}_{\tau}\right) \tau^c \; ,
\label{5.4}
\end{eqnarray}
where $(\alpha \cdot \beta) = \alpha^{}_1 \beta^{}_1 + \alpha^{}_2 \beta^{}_2 + \alpha^{}_3 \beta^{}_3 $ denotes the contraction of two triplets into a singlet, $y^{}_i$ and $y^{}_{\alpha}$ are dimensionless coefficients, and $\Lambda$ is the typical energy scale where the flavor-symmetry physics resides. In the literature, the ratios of the flavon VEVs to $\Lambda$ are usually assumed to be small so that the contributions of higher-dimension terms are suppressed. If the flavon fields possess the following VEV alignments
\begin{eqnarray}
&& \langle \phi^{}_1 \rangle  = v^{}_1 \left( \begin{array}{c} 1 \cr 0 \cr 1 \end{array} \right) \; , \hspace{0.5cm}
\langle \phi^{}_2 \rangle = v^{}_2 \left( \begin{array}{c} 1 \cr 1 \cr -1 \end{array} \right) \; ,  \hspace{0.5cm}
\langle \phi^{}_3 \rangle = v^{}_3 \left( \begin{array}{c} 0 \cr 1 \cr 1 \end{array} \right) \; ,  \nonumber \\
&& \langle \phi^{}_e \rangle = v^{}_e \left( \begin{array}{c} 1 \cr 0 \cr 0 \end{array} \right) \; , \hspace{0.5cm}
\langle \phi^{}_\mu \rangle = v^{}_\mu \left( \begin{array}{c} 0 \cr 1 \cr 0 \end{array} \right) \; , \hspace{0.5cm}
\langle \phi^{}_\tau \rangle = v^{}_\tau \left( \begin{array}{c} 0 \cr 0 \cr 1 \end{array} \right) \; ,
\label{5.5}
\end{eqnarray}
and $\langle \xi \rangle = M^{}_3$, then one arrives at a diagonal charge-lepton mass matrix and $M^{}_{\rm R}$ and an $M^{}_{\rm D}$ of the form in Eq.~(\ref{5.3}).

Then, following the idea of Ref.~\cite{littlest}, we justify the flavon VEV alignments in Eq.~(\ref{5.5}) by means of the F-term alignment mechanism \cite{F-term}. For this purpose, some driving fields $A$ are introduced, which carry an R charge of 2 and couple with the flavon fields linearly to form certain superpotential terms. In this way the minimization requirement of the potential energy $V(\phi) = \sum |\partial W/\partial A|^2$ leads to the constraint $\partial W/\partial A =0$ for the flavon VEVs. We note that the VEV alignments of $\phi^{}_\alpha$, $\phi^{}_2$ and $\phi^{}_3$ in Eq.~(\ref{5.5}) are the same as in Ref.~\cite{littlest} (see Eqs.~(13.2, 13.3) there), so they can be achieved in the same way as there. Hence one just needs to demonstrate that the VEV alignment of $\phi^{}_1$ in Eq.~(\ref{5.5}) can be naturally achieved. Since the VEV alignments of $\phi^{}_1$ and $\phi^{}_3$ only differ by a permutation of the first and second components, the former can be achieved in a way similar to the latter: on the one hand, the superpotential term $A^{}_{12} (\phi^{}_1 \cdot \phi^{}_2)$ where $A^{}_{12}$ is a driving field with the transformation properties $({\bf 1^\prime}, 1, 1)$ under $\rm S^{}_4 \times Z^{}_2 \times Z^{}_3$ will lead to the orthogonality
\begin{eqnarray}
\partial W/\partial A^{}_{12} = 0 = (\langle \phi^{}_1 \rangle \cdot \langle \phi^{}_2 \rangle) = \langle \phi^{}_1 \rangle^{}_1 \langle  \phi^{}_2 \rangle^{}_1 +  \langle \phi^{}_1 \rangle^{}_2 \langle  \phi^{}_2 \rangle^{}_2 +  \langle \phi^{}_1 \rangle^{}_3 \langle  \phi^{}_2 \rangle^{}_3 \;,
\label{5.6}
\end{eqnarray}
of the VEV alignments of $\phi^{}_1$ and $\phi^{}_2$.
On the other hand, the superpotential terms
$A^{}_1 (g^{}_1 \phi^{}_1 \phi^{}_1 + g^\prime_1 \xi^{}_1 \phi^{}_\mu )$
where $A^{}_1$ is a driving field with the transformation property $({\bf 3}, 1, 1)$ under $\rm S^{}_4 \times Z^{}_2 \times Z^{}_3$ and $\xi^{}_1$ has the transformation property $({\bf 1}, -1, \omega^2)$
will lead to the following constraint on the VEV alignment of $\phi^{}_1$:
\begin{eqnarray}
2 g^{}_1 \left( \begin{array}{c}
\langle \phi^{}_{1} \rangle^{}_2  \langle \phi^{}_{1} \rangle^{}_3 \cr
\langle \phi^{}_{1} \rangle^{}_3  \langle \phi^{}_{1} \rangle^{}_1 \cr
\langle \phi^{}_{1} \rangle^{}_1  \langle \phi^{}_{1} \rangle^{}_2 \end{array} \right)
+ g^\prime_1 \langle \xi^{}_1   \rangle
\left( \begin{array}{c}
\langle \phi^{}_{\mu} \rangle^{}_1 \cr
\langle \phi^{}_{\mu} \rangle^{}_2 \cr
\langle \phi^{}_{\mu} \rangle^{}_3 \end{array} \right)
= \left( \begin{array}{c} 0 \cr 0 \cr 0 \end{array} \right) \; .
\label{5.7}
\end{eqnarray}
Taking account of the VEV alignments of $\phi^{}_2$ and $\phi^{}_\mu$ in Eq.~(\ref{5.5}), the combination of Eqs.~(\ref{5.6}, \ref{5.7}) then yields $\langle \phi^{}_{1} \rangle \propto (1, 0, 1)^T$.

To justify the particular value of $\phi$, one needs to impose the CP symmetry (so that the coefficients are constrained to be real) and then break it in a particular way (so that a non-trivial CP phase can arise) \cite{cpsymmetry, LS}.
In the present model, as mentioned in the above, the $\rm Z^{}_3$ symmetry can help us fulfill this purpose: the superpotential term $A^{}_\xi ( \xi^3/\Lambda - M^2)$ where $A^{}_\xi$ is a singlet driving field and $M$ is a real (as constrained by the CP symmetry) mass parameter will lead to the constraint $\langle \xi \rangle^3/\Lambda - M^2 =0$ on the VEV of $\xi$, which can give $M^{}_3 = \langle \xi \rangle = e^{{\rm i} 2\pi/3} M$. After a phase redefinition of the $N^{}_3$ field, $M^{}_3$ can be made to be real again but the third column of $M^{}_{\rm D}$ will receive a common phase (i.e., $\phi$) of $-\pi/3$. A simple generalization of such an exercise can help us achieve $\phi = - \pi/n$ (with $n$ being an integer) with the help of a ${\rm Z}^{}_n$ symmetry.

Finally, we emphasize that the particular VEV alignments in Eq.~(\ref{5.5}) are associated with the S$^{}_4$ symmetry itself, but not necessarily associated with the F-term alignment mechanism which works in the SUSY framework where the reheating temperature above $\sim 10^9$ GeV might lead to the problem of gravitino overproduction \cite{gravitino}. The latter is just a tool that is commonly used in the literature to show the desired VEV alignments can be naturally realized. Alternatively, one can employ the D-term alignment mechanism \cite{d-term} to fulfill such a purpose, which is also applicable in the non-SUSY context where there would not be a gravitino problem.

\section{Impacts of the renomarlization group running effects}

Finally, in consideration of the huge gap between the seesaw scale where the texture of $M^{}_{\rm D}$ forms and leptogenesis takes place and the electroweak scale where the neutrino parameters are measured,
we give some discussions about the impacts of the renormalization group running effects on the texture of $M^{}_{\rm D}$ \cite{RGE} and leptogenesis \cite{giudice}.

In the SM framework, the Dirac neutrino mass matrix $M^{}_{\rm D}(\Lambda^{}_{\rm SS})$ at the seesaw scale is connected with its counterpart $M^{}_{\rm D}(\Lambda^{}_{\rm EW})$ at the electroweak scale through a relation as \cite{IRGE}
\begin{eqnarray}
M^{}_{\rm D} (\Lambda^{}_{\rm SS}) & = & I^{}_{0} \left( \begin{array}{ccc}
\vspace{0.1cm}
1-\Delta^{}_{e} &   &  \cr
 & 1 -\Delta^{}_{\mu} &  \cr
 &  &  1-\Delta^{}_{\tau} \cr
\end{array} \right)
M^{}_{\rm D} (\Lambda^{}_{\rm EW}) \;,
\label{5.1}
\end{eqnarray}
where
\begin{eqnarray}
I^{}_{0} & = & {\rm exp} \left( \frac{1}{32 \pi^2} \int^{\rm ln (\Lambda^{}_{SS}/\Lambda^{}_{\rm EW})}_{0} \left[ \lambda(t) - 3 g^2_2(t)  + 6 y^2_t(t)  \right] \ {\rm dt} \right) \;, \nonumber \\
\Delta^{}_{\alpha}  & = &  \frac{3}{32 \pi^2}\int^{\rm ln (\Lambda^{}_{SS}/\Lambda^{}_{\rm EW})}_{0} y^2_{\alpha}(t) \ {\rm dt} \;,
\label{5.2}
\end{eqnarray}
with $\lambda(t)$, $g^{}_2(t)$, $y^{}_t(t)$ and $y^{}_{\alpha}(t)$ standing respectively for the energy-scale-dependent Higgs quartic coupling, ${\rm SU(2)^{}_{L}}$ gauge
coupling, top-quark Yukawa coupling and charged-lepton Yukawa couplings. Qualitatively, $I^{}_0$ is just an overall rescaling factor for $M^{}_{\rm D}$ (i.e., only relevant for its overall scale but does not modify its texture), while $\Delta^{}_\alpha$ are potentially capable of modifying its texture.

Quantitatively, due to the smallness of $y^{}_\alpha$, $\Delta^{}_\alpha$ are negligibly small: $\Delta^{}_{\tau}$ is merely $\mathcal O(10^{-5})$, and $\Delta^{}_{e}$ and $\Delta^{}_{\mu}$ are much smaller. Even in the MSSM where $y^{2}_\tau = (1+ \tan^2{\beta}) m^2_\tau/v^2$ can be greatly enhanced by  large $\tan \beta$ values, one still has $\Delta^{}_{\tau} \lesssim 0.01 $ for a reasonable $\tan \beta$ value (e.g., $\lesssim 30$).
This means that the impacts of the renormalization group running effects on the texture of $M^{}_{\rm D}$ can be safely neglected. Therefore, although in the above study we have been confronting the considered textures of $M^{}_{\rm D}$ against the values of the neutrino parameters at the electroweak scale, the conclusions we have reached will hold equally well at the seesaw scale.

In comparison, one has $I^{}_0 \sim 1.15$, which lifts the overall scale of $M^{}_{\rm D} (\Lambda^{}_{\rm SS})$ by about $15\%$ as compared to $M^{}_{\rm D} (\Lambda^{}_{\rm EW})$.
Such a result has the following two impacts on leptogenesis which are in opposite directions: on the one hand, the CP asymmetry for the decays of the right-handed neutrinos gets enhanced by $I^2_0 \sim 1.23$ (see Eqs.~(\ref{3.11}, \ref{3.12})). On the other hand, the washout mass parameter also gets enhanced by $I^2_0$ (see Eq.~(\ref{3.13})), making the washout effects more efficient. As is known, the washout mass parameter indicated by neutrino oscillations lies in the strong washout regime where the efficiency factor is roughly inversely proportional to it \cite{review}. Consequently, the efficiency factor roughly gets suppressed by $I^2_0$. Altogether, for the final baryon asymmetry, the suppression of the efficiency factor offsets the enhancement of the CP asymmetry to a large degree. This makes the impacts of the renormalization group running effects on leptogenesis acceptably small (within the 10 percent level).

\section{Summary}

In summary, due to their simple structure and phenomenologically-appealing consequences, the trimaximal mixings have attracted a lot of attention. In this paper, in the basis of $M^{}_{\rm R}$ being diagonal, we have explored the simplified textures of $M^{}_{\rm D}$ that can naturally yield these mixings and their consequences for the neutrino parameters and leptogenesis.

We have first formulated the generic textures of $M^{}_{\rm D}$ that can naturally yield the trimaximal mixings (see Eq.~(\ref{2.5})) and discussed how to realize them by slightly modifying the flavor-symmetry models for realizing the TBM mixing. We have then examined if their parameters can be further reduced, giving more simplified textures of them. Our analysis has been restricted to the simple but instructive scenario that three elements in the same column of $M^{}_{\rm D}$ share a common phase. Furthermore, for the TM1 (TM2) mixing, only the phase difference between the second and third (first and third) columns is responsible for $\delta$ and leptogenesis, while the phase of the first (second) column only contributes to $\rho$ ($\sigma$) additively. Therefore, without loss of generality, our analysis has been further restricted to the scenario that there is only one phase parameter $\phi$ (i.e., the third-column phase), in which case $M^{}_{\rm D}$ can be conveniently reexpressed as in Eqs.~(\ref{3.1}, \ref{4.1}).

It should be noted that the equality between $x$ and $y$ is denied, which would otherwise lead to the unacceptable $\delta =0$. For the TM1 (TM2) mixing, the results of $(x, y) = (y^{}_0, x^{}_0)$ can be obtained from those of $(x, y) = (x^{}_0, y^{}_0)$ by making the replacements $\phi \to -\phi$ and $\rho \to \rho + \phi$ ($\sigma \to \sigma + \phi$). So we have just considered the $x<y$ cases. Furthermore, the results of $(x, y) = (-y^{}_0, -x^{}_0)$ can be obtained from those of $(x, y) = (x^{}_0, y^{}_0)$ by making the replacements $\rho \to -(\rho + \phi)$ ($\sigma \to -(\sigma + \phi)$), $\sigma \to - \sigma$ ($\rho \to - \rho$), $\delta \to \pi - \delta$ and $\Delta s^{2}_{23} \to - \Delta s^{2}_{23}$.

From the simplicity viewpoint, we aim to explore the simplified textures of $M^{}_{\rm D}$ that can naturally yield the trimaximal mixings where there are some vanishing or equal elements.
Such textures of $M^{}_{\rm D}$ correspond to some particular values of $x$ and $y$ (see Tables~\ref{Table2} and \ref{Table5}). But our discussions have been restricted to the textures of $M^{}_{\rm D}$ that can find a simple symmetry justification.
The phenomenologically-viable particular $(x, y)$ combinations and the allowed ranges of $m^{}_l$, $\phi$, $\delta$, $\rho$ and $\sigma$ are listed in Tables~\ref{Table3} and \ref{Table6}.
On the basis of these particular $(x, y)$ combinations, we have further examined if $\phi$ can also take some particular value. The phenomenologically-viable particular $(x, y, \phi)$ combinations and their predictions for the neutrino parameters at $\chi^2_{\rm min}$ are listed in Tables~\ref{Table4} and \ref{Table7}. Finally, the consequences of these particular $(x, y, \phi)$ combinations for leptogenesis have been studied.
Because of the special form of $M^{}_{\rm D}$, for the TM1 (TM2) mixing, the final baryon asymmetry can only be owing to $N^{}_2$ ($N^{}_1$) or $N^{}_3$. But the washout effects from $N^{}_1$ ($N^{}_2$) may be non-negligible when it is the lightest one. Taking account of the interplay between the right-handed neutrino mass spectrum and the flavor effects, there are several possible scenarios for leptogenesis. For these different scenarios, the values of $M^{}_2$ ($M^{}_1$) or $M^{}_3$ for leptogenesis to be viable are calculated and listed in Tables~\ref{Table42} and \ref{Table72}.

Then, a concrete S$^{}_4$-flavor-symmetry model that can realize one representation of the obtained simplified textures of $M^{}_{\rm D}$ is given. And the F-term alignment mechanism and CP symmetry are invoked to justify the particular VEV alignments of the flavon fields and the non-trivial value of the CP phase.

Finally, the impacts of the renormalization group running effects on the texture of $M^{}_{\rm D}$ and leptogenesis have also been discussed. It is found that the impacts of the renormalization group running effects on the texture of $M^{}_{\rm D}$ can be safely neglected. And the impacts of the renormalization group running effects on leptogenesis are also acceptably small (within the 10 percent level).

In summary, our work, in the general seesaw framework, provides a complete study of the simplest textures of $M^{}_{\rm D}$ (which are motivated from the simplicity viewpoint and can find a simple symmetry justification) that can naturally yield the trimaximal (including both the TM1 and TM2) mixings, and their consequences for the neutrino parameters and leptogenesis. Since they only contain four real parameters (see, e.g., Eq.~(53)), they are very restrictive and highly predictive. Their predictions for the neutrino parameters can be tested or ruled out by future precision measurements. Furthermore, since there is only one CP phase, a direct link between the CP violating effects at low energies and leptogenesis can be established.

\vspace{0.5cm}

\underline{Acknowledgments} \hspace{0.2cm}
This work is supported in part by the National Natural Science Foundation of China under grant Nos. 11605081 and 12047570, and the Natural Science Foundation of the Liaoning Scientific Committee under grant NO. 2019-ZD-0473.

\end{document}